\documentclass[reqno,10pt,centertags]{amsart}
\usepackage{amsmath,amsthm,amscd,amssymb,latexsym,upref}
\date{\today}

\newcommand{\bbD}{{\mathbb{D}}}

\newcommand{\bbR}{{\mathbb{R}}}

\newcommand{\bbC}{{\mathbb{C}}}

\newcommand{\bbT}{{\mathbb{T}}}
\newcommand{\mcB}{{\mathcal{B}}}
\newcommand{\fB}{{\mathfrak{B}}}
\newcommand{\fA}{{\mathfrak{A}}}

\renewcommand{\Re}{\text{\rm Re\, }}
\renewcommand{\Im}{\text{\rm Im\, }}
\newcommand{\scp}{{\{s_+,\nu_+\}}}
\newcommand{\scm}{{\{s_-,\nu_-\}}}
\newcommand{\Ms}{\langle s_+ \rangle_{I}}

\newcommand{\br}{{\mathbb{R}}}

\newcommand{\bz}{{\mathbb{Z}}}
\newcommand{\bc}{{\mathbb{C}}}

\renewcommand{\l}{\lambda}

\newcommand{\ep}{\varepsilon}
\renewcommand{\d}{\delta}

\newcommand{\re}{\text{\rm Re\,}}
\newcommand{\im}{\text{\rm Im\,}}
\newcommand{\scpn}{{\{s_+,\nu_{+ N}\}}}
\newcommand{\scpx}{{\{s_+e^{2i\l x},\, \nu_+e^{2i\l x}\}}}
\newcommand{\scpxn}{{\{s_+e^{2i\l x},\, \nu_{+ N} e^{2i\l x}\}}}

\newcommand{\scpxne}{{\{s_{+,\ep}e^{2i\l x},\, \nu_{+N,\ep}e^{2i\l x}\}}}
\newcommand{\scpxx}{{\{s_+e^{-2i\l x},\, \nu_+e^{-2i\l x}\}}}

\newcommand{\scpxxbn}{{\{s_+e^{-2i\l x}(b_{\l_0}b_{-\bar\l_0})^{-1},\,
\nu_{+ N}e^{-2i\l x}(b_{\l_0}b_{-\bar\l_0})^{-1}\}}}
\newcommand{\scpbn}{{\{s_+(b_{\l_0}b_{-\bar\l_0})^n,\nu_+(b_{\l_0}b_{-\bar\l_0})
^n\}}}

\newcommand{\scmb}{{\{s_-(b_{\l_0}b_{-\bar\l_0})^{-1},\, 
\nu_-(b_{\l_0}b_{-\bar\l_0})^{-1}\}}}
\newcommand{\scmx}{{\{s_- e^{2i\l x},\, \nu_- e^{2i\l x}\}}}
\newcommand{\scmxx}{{\{s_- e^{-2i\l x}\, \nu_- e^{-2i\l x}\}}}
\newcommand{\scmxn}{{\{s_- e^{2i\l x},\, \nu_{- N}e^{2i\l x}\}}}
\newcommand{\scmxb}{{\{s_- e^{2i\l x} (b_{\l_0}b_{-\bar\l_0})^{-1},\, \nu_- 
e^{2i\l x}(b_{\l_0}b_{-\bar\l_0})^{-1}\}}}
\newcommand{\scmxxb}{{\{s_- e^{-2i\l x} (b_{\l_0}b_{-\bar\l_0})^{-1},\, \nu_- 
e^{-2i\l x}(b_{\l_0}b_{-\bar\l_0})^{-1}\}}}
\newcommand{\scmxxbn}{{\{s_- e^{-2i\l x}(b_{\l_0}b_{-\bar\l_0})^{-1},\, \nu_{- 
N} 
e^{-2i\l x}(b_{\l_0}b_{-\bar\l_0})^{-1}\}}}

\newcommand{\scmxbne}{{\{s_{-,\ep} e^{-2i\l x} (b_{\l_0}b_{-\bar\l_0})^{-1},\,
\nu_{-N,\ep} e^{-2i\l x}(b_{\l_0}b_{-\bar\l_0})^{-1}\}}}

\newcommand{\lt}{\left}
\newcommand{\rt}{\right}
\newcommand{\ti}{\tilde}

\allowdisplaybreaks
\numberwithin{equation}{section}
\newtheorem{theorem}{Theorem}[section]

\newtheorem{lemma}[theorem]{Lemma}
\newtheorem{corollary}[theorem]{Corollary}

\theoremstyle{definition}
\newtheorem{definition}[theorem]{Definition}
\newtheorem{remark}[theorem]{Remark}

\begin{document}

\title[]{Inverse scattering
problem for a special class of canonical 
systems and non-linear Fourier integral. Part I: asymptotics of eigenfunctions} 
\author[]{S. Kupin, F. Peherstorfer, A. Volberg, 
and P. Yuditskii}

\address{CMI, Universit\'e de Provence, 39, rue Joliot-Curie, 13453 Marseille 
Cedex 13, France}
\email{kupin@cmi.univ-mrs.fr}

\address{Institute for Analysis, Johannes Kepler University Linz,
A-4040 Linz, Austria}
\email{Franz.Peherstorfer@jk.uni-linz.ac.at}

\address{Department of Mathematics, Michigan State University, East Lansing MI 
48824, USA}
\email{volberg@math.msu.edu}

\address{Department of Mathematics, Bar Ilan University, Ramat Gan 52900, 
Israel}
\email{yuditski@mailhost.cs.biu.ac.il}

\thanks{Partially supported by NSF grant DMS-0200713 and
the Austrian Science Found FWF, project number P16390-N04.} 
\date{\today}

\begin{abstract}
An original approach to the inverse scattering for Jacobi matrices was recently 
suggested in \cite{VYu}. The authors considered quite sophisticated spectral 
sets (including Cantor sets of positive Lebesgue measure), however they did not 
take into account the mass point spectrum. 
This paper follows similar lines for the conti\-nuous setting with an absolutely 
continuous spectrum on the half-axis and a pure point spectrum on the negative 
half-axis satisfying the Blaschke condition.  This leads us to the solution of 
the inverse scattering problem for a class of canonical systems that generalizes 
the case of Sturm-Liouville (Schr\"odinger) operator.
\end{abstract}

\maketitle

\section{Faddeev-Marchenko space in Szeg\H o/Blaschke setting}
\label{s1}
One of the important aspects of the spectral theory of differential operators is 
the scattering theory \cite{rs1,rs2} and, in particular, the inverse scattering 
\cite{Mar}. An original approach to the inverse scattering was recently 
suggested in \cite{VYu}. The paper focused on classical Jacobi matrices and 
connections between the scattering and properties of a special Hilbert 
transform. 

In this paper, we carry out the plan of \cite{VYu} in the continuous situation. 
Compared with \cite{VYu}, a completely new feature is that the scattering data 
incorporate the pure point spectrum with infinitely many mass points. 
Of course, this is a natural and important step in the developing the theory. 
The discussion leads us to the solution of the inverse scattering problem for a 
class of canonical systems that include the
Sturm-Liouville (Schr\"odinger) equations. At present, though, we are unable to 
characterize the scattering data corresponding to the last important 
special case.

This part of the work is mainly devoted to the asymptotic behavior
of certain reproducing kernels (the generalized eigenfunctions). It is organized 
as follows. Section \ref{s1} contains definitions, some 
general facts and formulations of results on asymptotics. The asymptotic 
properties of reproducing kernels from certain model spaces are studied in 
Section \ref{s2}. Special operator nodes arising from our construction are 
discussed in Sections \ref{s3} and \ref{s4}. One of the nodes generates a 
canonical system we are interested in. Its properties and connections to the de 
Branges spaces of entire functions \cite{dB} are also in Section \ref{s4}. The 
Sturm-Liouville (Schr\"odinger) equations are considered in Section \ref{s5}. An 
example is given in the first appendix (Section \ref{s6}). The second appendix 
(Section  \ref{s7})
relates the whole construction to the matrix $A_2$ Hunt-Muckenhoupt-Wheeden 
condition. 

We define the $L^2$-norm on the real axis as
\begin{equation}\label{1.9d}
||f||^2=\frac{1}{2\pi}\int_{\bbR}|f(\lambda)|^2\,d\lambda,
\end{equation}
so that the reproducing kernel of the $H^2$ subspace is of the form
$k(\lambda,\lambda_0)=
\frac{i}{\lambda-\bar\lambda_0}$.

The section  ``Inverse
scattering problem on the real axis" in \cite[Chap. 3,
Sect. 5]{Mar} begins with a Sturm-Liouville operator
\begin{equation}\label{2.9d}
-y''+q(x)y=\lambda^2 y,\quad x\in\bbR,
\end{equation}
with the potential $q$ satisfying the a priori condition
\begin{equation}\label{3.9d}
\int_{\bbR}(1+|x|)|q(x)|\,dx<\infty.
\end{equation}
To such an operator one associates so called scattering data
\begin{equation}\label{4.9d}
\{s_+,\nu_+\},
\end{equation}
where $s_+$ is a contractive function on the real axis,
$|s_+(\lambda)|\le 1,\ \lambda\in\bbR$,
possessing
certain properties and $\nu_+$ is a discrete measure, in fact, 
supported on a finite number of points $\Lambda=\{\lambda_k\}$
of the imaginary axis, $\frac{\lambda_k}i>0$.

We proceed in the opposite direction starting from the scattering data 
$\{s_+,\nu_+\}$ and going to the potential $q$. The key point of the 
construction is that we assume that the scattering data  \eqref{4.9d} satisfy 
only very natural (and minimal) conditions from the point of view of the 
function theory. Namely, we 
suppose that:
\begin{itemize}
\item[--]
a symmetric on the real axis function $s_+$, $s_+(\lambda)=
\overline{s_+(-\bar\lambda)}$, satisfies the Szeg\H o condition
\begin{equation}\label{5.9d}
\int_{\bbR}\frac{\log(1-|s_+(\lambda)|^2)}{1+\lambda^2}d\lambda>-\infty,
\end{equation}
\item[--] the support $\Lambda$ of
a discrete measure $\nu_+=\sum_k \nu_+(\l_k)\d_{\l_k}$ satisfies the Blaschke 
condition
\begin{equation}\label{6.9d}
\sum_{(\lambda_k/i)\le 1}\frac{\lambda_k}{i}<\infty,\quad
\sum_{(\lambda_k/i)> 1}\frac{i}{\lambda_k}<\infty.
\end{equation}
\end{itemize}
Let us point out that we  did not even assume that the measure
$\nu_+$ is finite.

Our plan is to show that already in this case one 
can associate a certain differential operator of the second order to the given 
spectral data and then one can prove several specification theorems.

\begin{definition}
An element $f$ of the space $L^2_{\scp}$
is a function on $\bbR\cup \Lambda$ such that
\begin{equation}\label{7.9d}
\begin{split}
||f||^2_{\scp}=& \sum_{\lambda_k\in\Lambda}|f(\lambda_k)|^2
\nu_+(\lambda_k)\\
               +&\frac 1 {4\pi}
               \int_{\bbR}\begin{bmatrix}
               \overline{f(\lambda)}& 
               \overline{-f(-\bar\lambda)}
               \end{bmatrix}
               \begin{bmatrix}
               1&  
               \overline{s_+(\lambda)}\\
               s_+(\lambda)& 1
               \end{bmatrix}
               \begin{bmatrix}
               {f(\lambda)}\\
               {-f(-\bar\lambda)}
               \end{bmatrix}d\lambda
               \end{split}
\end{equation}
is finite.
\end{definition}

Using \eqref{5.9d} and \eqref{6.9d} 
we define the outer in the upper half-plane
function $s_e$ as
\begin{equation}\label{8.9d}
|s_e(\lambda)|^2+|s_+(\lambda)|^2=1\ \text{a.e. on}\ \bbR,
\end{equation}
and the Blaschke product
\begin{equation}\label{9.9d}
B(\lambda)=\prod_k b_{\l_k}(\l),
\end{equation}
where $b_{\l_k}(\l)=\frac{\lambda-\lambda_k} {\lambda-\bar\lambda_k}$ if 
$\l_k/i\le 1$ and 
$b_{\l_k}(\l)=-\frac{\lambda-\lambda_k} {\lambda-\bar\lambda_k}$ if $\l_k/i>1$.
We also put 
\begin{equation}\label{1.10d}
S(\lambda)=\begin{bmatrix}
s_-&s\\
s& s_+
\end{bmatrix}(\lambda),\ \lambda\in\bbR,
\end{equation}
where
\begin{equation}\label{2.10d}
s:=\frac{s_e}B\quad\text{and}\quad
s_-:=-\frac s{\bar s}\bar s_+.
\end{equation}
The matrix function $S$ possesses two fundamental properties:
$S^*(-\bar\lambda)=S(\lambda)$, and it is unitary-valued.
The third property is analyticity of the entry $s$, which has
analytic continuation to the upper half-plane as a function of bounded
characteristic with a specific nature, that is, it is a ratio of an outer
function and a Blaschke product.

The measure $\nu_-$ is defined through $s_-$ and $\nu_+$ by
\begin{equation}\label{3.10d}
\frac 1{\nu_+(\lambda_k)}\frac 1{\nu_-(\lambda_k)}=
\left|\left(\frac 1{s}\right)'(\lambda_k)\right|^2.
\end{equation}
A reason for these and the following definitions will be clarified in a
moment.

Set
\begin{equation}\label{4.10d}
\begin{split}
\begin{bmatrix}
sf^+\\ s f^-
\end{bmatrix}(\lambda)=&
\begin{bmatrix}
s&0\\ s_+&1
\end{bmatrix}(\lambda)
\begin{bmatrix}
f^+(\lambda)\\  -f^+(-\bar\lambda)
\end{bmatrix}\\
=&
\begin{bmatrix}
1&s_-\\ 0&s
\end{bmatrix}(\lambda)
\begin{bmatrix}
-f^-(-\bar\lambda)\\  f^-(\lambda)
\end{bmatrix}
\end{split}
\end{equation}
for $\lambda\in\bbR$ and
\begin{equation}\label{5.10d}
f^{-}(\lambda_k)=-i\left(\frac 1 s\right)'(\lambda_k)
{\nu_+(\lambda_k)}{f^+(\lambda_k)}
\end{equation}
for $\lambda_k\in\Lambda$. It is evident that in this way we define
a unitary map from $L^2_{\scp}$ to $L^2_{\scm}$. In fact, due to
\eqref{4.10d}
\begin{equation}\label{1.22d}
\frac 1 {4\pi}
               \int_{\bbR}\begin{bmatrix}
               \overline{f(\lambda)}& 
               \overline{-f(-\bar\lambda)}
               \end{bmatrix}
               \begin{bmatrix}
               1&  
               \overline{s_+(\lambda)}\\
               s_+(\lambda)& 1
               \end{bmatrix}
               \begin{bmatrix}
               {f(\lambda)}\\
               {-f(-\bar\lambda)}
               \end{bmatrix}d\lambda=
               \frac{||sf^+||^2+||sf^-||^2}2,
\end{equation}
where we have the standard $L^2$-norm on $\bbR$ in the RHS of the equality.
The key point is that relations \eqref{4.10d}, \eqref{4.10d} not only define a 
duality between these two spaces but, what is more important,  a duality between 
corresponding Hardy subspaces.

Actually we give two versions of  definitions of Hardy subspaces
(in general, {\it they are not equivalent},
see an example in  Section \ref{s6}). By the first one,
$H^2_{\scp}$ is basically the closure of $H^\infty$
with respect to the given norm \eqref{7.9d}. More precisely,
let $B_{N_1,N_2}=\prod_{k=N_1+1}^{N_2} b_{\l_k}$ and $\mcB=\{B_{N,\infty}\}$. 
Saying it differently, $B'\in \mcB$ if and only if $B'$ is a divisor of $B$ such 
that
$B/B'$ is a finite Blaschke product. Then 
\begin{equation}\label{10.9d}
f=B_{N,\infty} g,\quad g\in H^\infty,
\end{equation}
belongs to $L^2_{\scp}$. By $H^2_{\scp}$ we denote  the closure
in $L^2_{\scp}$ of functions of the form \eqref{10.9d}.
Let us point out that every element $f$ of $H^2_{\scp}$
is such that $s_ef$ belongs to the standard $H^2$,
see \eqref{1.22d}. Therefore,
in fact, $f(\lambda)$ has an analytic continuation from the real
axis to the upper half-plane. Moreover, the value $f(\l)$
obtained by this  continuation,  and 
$f(\lambda_k)$ which is defined for
all $\lambda_k\in\Lambda$,
since $f$ is a function from
$L^2_{\scp}$,
still perfectly coincide. 

The second space also consists of functions from
$L^2_{\scp}$ having an analytic continuation to the upper
half-plane.
\begin{definition}\label{d11} A function $f\in L^2_{\scp}$ belongs to
$\hat H^2_{\scp}$ if $g(\lambda):=(s_e f)(\lambda)$,
$\lambda\in \bbR$, belongs to the standard $H^2$ and
$$
f(\lambda_k)=\left(\frac{g}{s_e}\right)(\lambda_k),
\ \lambda_k\in\Lambda,
$$
where in the RHS $g$ and $s_e$ are defined by  their analytic
continuation to the upper half-plane.
\end{definition}

It turns out that spaces $H^2_\scp$ and $\hat H^2_\scp$ are dual in a certain 
sense.

\begin{theorem}\label{t1.3}
Let $f^+\in L^2_{\scp}\ominus H^2_{\scp}$
and let $f^-\in L^2_{\scm}$ be defined by \eqref{4.10d},
\eqref{5.10d}. Then $f^-\in
\hat H^2_{\scm}$. In short, we write
\begin{equation}\label{0.17d}
(\hat H^2_{\scm})^+= L^2_{\scp}\ominus H^2_{\scp}.
\end{equation}
\end{theorem}
\begin{proof}
We notice that $f^+\in L^2_{\scp}$ implies
$$
(sf^-)(\lambda)=s_+(\lambda)f^+(\lambda)-f^+(-\bar\lambda)\in L^2,
\ \lambda\in\bbR.
$$
Since
$$
\langle f^+, Bh \rangle_{\scp}=
\langle
s_+(\lambda)f^+(\lambda)-f^+(-\bar\lambda),
-B(-\bar\lambda)h(-\bar\lambda)
\rangle,\ h\in H^2,
$$
it follows from
$f^+\in L^2_{\scp}\ominus H^2_{\scp}$ that
$$
(s_ef^{-})(\lambda)=g(\lambda)
:=B(\lambda)(s_+(\lambda)f^+(\lambda)-f^+(-\bar\lambda))\in
H^2.
$$
Now we calculate the scalar product
\begin{equation*}
\begin{split}
\langle f^+, \frac{iB(\lambda)}{\lambda-\lambda_k} \rangle_{\scp}=&
f^+(\lambda_k)\overline{iB'(\lambda_k)}\nu_+(\lambda_k)
+\langle
s_ef^-,\frac{i}{\lambda-\bar\lambda_k}
\rangle\\=&
f^+(\lambda_k)iB'(\lambda_k)\nu_+(\lambda_k)
+
g(\lambda_k)=0.
\end{split}
\end{equation*}
Therefore, by \eqref{5.10d} we get
$$
f^-(\lambda_k)=\left(\frac{g}{s_e}\right)(\lambda_k),
\ \lambda_k\in\Lambda.
$$
\end{proof}

Both $H^2_{\scp}$ and $\hat H^2_{\scp}$ are spaces
of analytic in the upper half-plane functions, so they have reproducing 
kernels.
For $\mu\in\bc_+$, we denote them by 
$$
k_{\scp}(\l,\mu)=k_{\scp}(.,\mu),\quad \hat k_{\scp}(\lambda,\mu)=\hat 
k_\scp(.,\mu).
$$
Recall also that
$$
k(\l,\mu)=k(.,\mu)=\frac{i}{\l-\bar\mu}
$$
is the reproducing kernel of the standard Hardy space $H^2$.
The first step is to prove asymptotics for the families $\{e^{i\l x}k_{\{s_\pm 
e^{2i\l x}, \nu\pm e^{2i\l x}\}}(.,\l_0)\}_{x\in \br}$ and  $\{e^{i\l x}$ $\hat 
k_{\{s_\pm e^{2i\l x}, \nu\pm e^{2i\l x}\}}(.,\l_0)\}_{x\in \br}$ with $\l_0\in 
\bc_+$.

\begin{theorem}\label{t1} The following relations hold true:
\begin{itemize}
\item[\it i)] on $\br$, 
\begin{eqnarray}
s \lt(e^{i\l x} k_\scpx (.,\l_0)\rt)&=&s e^{i\l x} k(.,\l_0)+o(1), 
\label{ee01}\\
s\lt(e^{i\l x} \hat k_\scpx (.,\l_0)\rt)&=&s e^{i\l x} k(.,\l_0)+o(1) \nonumber
\end{eqnarray}
as $x\to +\infty$. Moreover,
\begin{eqnarray}
s(-\bar\l_0)s\lt(e^{i\l x} k_\scpx (.,\l_0)\rt)&=&e^{i\l x}k(.,\l_0) 
\label{ee02}\\
&+&s_- e^{-i\l x}k(.,-\l_0)+o(1), \nonumber\\
s(-\bar\l_0)s\lt(e^{i\l x} \hat k_\scpx (.,\l_0)\rt)&=&e^{i\l 
x}k(.,\l_0)\nonumber\\
&+&s_- e^{-i\l x}k(.,-\l_0)+o(1) \nonumber
\end{eqnarray}
as  $x\to -\infty$ (of course, everything is in $L^2$-sense).
\item[\it ii)] on $\Lambda$,
\begin{equation}\label{ee03}
e^{i\l x} k_\scpx (.,\l_0)=o(1),\quad e^{i\l x} \hat k_\scpx (.,\l_0)=o(1)
\end{equation}
in $L^2_{\nu_+}$-sense as $x\to+\infty$. Furthermore, for a $\l_k\in\Lambda$
\begin{eqnarray}\label{ee04}
\lim_{x\to-\infty} e^{-2\im\l_k x} k_\scpx(\l_k,\l_k)&=&\frac 1{\nu_+(\l_k)},\\
\lim_{x\to-\infty} e^{-2\im\l_k x} \hat k_\scpx(\l_k,\l_k)&=&\frac 
1{\nu_+(\l_k)}\nonumber
\end{eqnarray}
\end{itemize}
\end{theorem}
It goes without saying that relations \eqref{ee01}--\eqref{ee03} correspond to 
scattering ``from $+\infty$ to $-\infty$''; compare these formulas to (0.8), 
(0.25) from \cite{VYu}. Scattering in the inverse direction 
(``from $-\infty$ to $+\infty$'') is described similarly. We give the formulas 
for the family $\{e^{-i\l x} k_\scmxx(.,\l_0)\}$ only; asymptotics for 
$\{e^{-i\l 
x}\hat k_\scmxx$ $(.,\l_0)\}$ are the same.

\begin{corollary}\label{c11} We have
\begin{eqnarray*}
s \lt(e^{-i\l x} k_\scmxx(.,\l_0)\rt)&=&s e^{-i\l x} k(.,\l_0)+o(1),\qquad 
x\to-\infty, \\
s(-\bar\l_0)s\lt(e^{-i\l x} k_\scmxx(.,\l_0)\rt)&=&e^{-i\l x}k(.,\l_0) \\
&+&s_+ e^{i\l x}k(.,-\l_0)+o(1),\quad x\to+\infty
\end{eqnarray*}
in $L^2$-sense on the real line. As for $\Lambda$,
\begin{equation*}
e^{-i\l x} k_\scmxx(.,\l_0)=o(1) 
\end{equation*}
in $L^2_{\nu_-}$-sense as $x\to-\infty$. As before, for a $\l_k\in\Lambda$,
\begin{equation*}
\lim_{x\to+\infty} e^{2\im\l_k x} k_\scmxx(\l_k,\l_k)=\frac 1{\nu_-(\l_k)}.
\end{equation*}
\end{corollary}
We set
$$
\ti k(.,\l_0)=
\lt\{
\begin{array}{ll}
k(.,\l_0),& \l\in\br,\\
0,& \l\in\Lambda.
\end{array}\rt.
$$
Theorem \ref{t1} follows immediately from the following result.
\begin{theorem}\label{t2} The following relations hold true:
\begin{eqnarray}
\lim_{x\to+\infty} ||e^{i\l x} k_\scpx(.,\l_0)-e^{i\l x}\ti 
k(.,\l_0)||_\scp&=&0, 
\label{ee41}\\
\lim_{x\to+\infty} ||e^{i\l x} \hat k_\scpx(.,\l_0)-e^{i\l x}\ti 
k(.,\l_0)||_\scp&=&0. \label{ee4}
\end{eqnarray}
\end{theorem}
The proof of this theorem is the main purpose of Section \ref{s2}.  

\begin{remark}\label{rk1} Relations \eqref{ee01}--\eqref{ee03} follow at once 
from Theorem \ref{t2}.
\end{remark}

Indeed, let us have a look at \eqref{ee41}. Recalling that $\ti k(.,\l_0)=0$ on 
$\Lambda$, we see
\begin{eqnarray*}
||e^{i\l x} k_\scpx(.,\l_0)&-&e^{i\l x}\ti k(.,\l_0)||^2_\scp\\
&=&||e^{i\l x} k_\scpx(.,\l_0)-e^{i\l x} k(.,\l_0)||^2_{s_+}\\
&+&||e^{i\l x} k_\scpx(.,\l_0)||^2_{\nu_+}\to 0,
\end{eqnarray*}
as $x\to +\infty$, so the first relation in \eqref{ee03} is proved. Then we 
notice that
$$
\lt[\begin{array}{cc}
1&\bar s_+\\
s_+& 1
\end{array}\rt]=
\lt[\begin{array}{cc}
|s|^2& 0\\
0& 0
\end{array}\rt]
+
\lt[\begin{array}{c}
\bar s_+\\
1\end{array}\rt]
\lt[\begin{array}{cc}
s_+&1\end{array}\rt].
$$
This implies that the first summand on the right-hand side of the above equality 
is
\begin{eqnarray*}
||\dots||^2_{s_+}&=&||s\lt(e^{i\l x} k_\scpx(.,\l_0)-e^{i\l x} 
k(.,\l_0)\rt)||^2_2\\
&+&
||s\lt(e^{i\l x} k_\scpx(.,\l_0)\rt)^-_{s_-}-\lt(e^{-i\l x}k(.,-\l_0)+s_+ e^{i\l 
x}k(.,\l_0)\rt)||^2_2.
\end{eqnarray*}
The presence of the first term on the right-hand side shows that we are done 
with \eqref{ee01}.
To deal with $(e^{i\l x} k_\scpx(.,\l_0))^-_{s_-}$, we use Lemma \ref{l2.1f5} 
and its corollary saying
$$
\lim_{x\to+\infty} k_\scmxxb(-\bar\l_0,-\bar\l_0)=\frac 1{2\im\l_0 |s(\l_0)|^2} 
$$
(see also Lemma \ref{l6}).
Hence, we come to
\begin{eqnarray*}
s(\l_0)s\lt(e^{-i\l x}\hat k_\scmxxb\rt)&=&e^{-i\l x}k(.,-\bar\l_0)\\
&+&s_+(b_{\l_0}b_{-\bar\l_0}) e^{i\l x} k(.,\bar\l_0)+o(1)
\end{eqnarray*}
as $x\to+\infty$. This is the second relation in \eqref{ee02} up to changes 
$x\mapsto -x,\l_0\mapsto -\bar\l_0, s_-(b_{\l_0}b_{-\bar\l_0})^{-1}\mapsto s_+$ 
and $s_+(b_{\l_0}b_{-\bar\l_0})\mapsto s_-$.

Equalities \eqref{ee04} are proved in Corollary \ref{c12}.

\section{Asymptotics of reproducing kernels}\label{s2}

\subsection{Definitions and notation}
In this subsection, we prove several propositions concerning 
special properties of the reproducing kernels introduced in Section \ref{s1}.

For $\l_0\in\bc_+$, let
\begin{equation*}
K_\scp(. ,\l_0)=\frac{k_\scp(. ,\l_0)}{\sqrt{k_\scp(\l_0,\l_0)}},\quad
\hat K_\scp(. ,\l_0)=\frac{\hat k_\scp(. ,\l_0)}{\sqrt{\hat k_\scp(\l_0,\l_0)}}
\end{equation*}
be their normalized versions. It is also convenient to put
$$
K(\l,\l_0)=\frac{k(\l,\l_0)}{||k(.,\l_0)||}=
\frac{i\, (2\im\l_0)^{1/2}}{\l-\bar\l_0}.
$$

For a fixed $x\in\bbR$ we define $H^2_{\scp}(x)$ as the closure
of the functions
\begin{equation}\label{1.16d}
f(\lambda)=B_{N,\infty}(\lambda) g(\lambda) e^{i\lambda x},\quad 
g\in H^\infty,\ B_{N,\infty}\in\mcB.
\end{equation}
In particular, $H^2_{\scp}=H^2_{\scp}(0)$. In the similar way
we define the set of spaces $\hat H^2_{\scp}(x)$, so that
$\hat H^2_{\scp}$ is related to $x=0$.

It is easy to see that 
$$
H^2_\scp(x)=e^{i\l x}H^2_\scpx,\quad
\hat H^2_\scp(x)=e^{i\l x}\hat H^2_\scpx,
$$
and 
\begin{eqnarray*}
k_\scp(\l,\l_0;x)&=&e^{ix(\l-\bar\l_0)}k_\scpx(\l,\l_0),\quad\\
\hat k_\scp(\l,\l_0;x)&=&e^{ix(\l-\bar\l_0)}\hat k_\scpx(\l,\l_0)
\end{eqnarray*}
are the reproducing kernels of these spaces, respectively. We also have their 
normalized versions
\begin{eqnarray*}
K_\scp(.,\l_0;x)&=&e^{ix(\l-\re\l_0)}K_\scpx(.,\l_0),\quad\\
\hat K_\scp(.,\l_0;x)&=&e^{ix(\l-\re\l_0)}\hat K_\scpx(.,\l_0).
\end{eqnarray*}

This section is mainly devoted to the proof of asymptotic formulas for both 
types of kernels as
$x\to+\infty$.

\subsection{Some special properties of the reproducing kernels}

The following lemma is trivial but probably 
the notations are slightly confusing. We belive that the diagram below will help
to avoid misunderstanding: $\pm$-mappings 
$ L^2_{\scp}\stackrel{\pm}{\longleftrightarrow} L^2_{\scm}$,
given by
\eqref{4.10d},
\eqref{5.10d}, actually depend on the scattering data
$\{s_\pm,\nu_\pm\}$, although we do not indicate
this dependence explicitly in most cases.

\begin{lemma}\label{l1.4} Let $w(\lambda)$ be an inner 
meromorphic function in the
upper half-plane such that $w(\lambda_k)\not=0$,
$w(\lambda_k)\not=\infty$
for all $\lambda_k\in\Lambda$. 
Put $w_*(\lambda):=\overline{w(-\bar\lambda)}$.
The following diagram
is commutative
\begin{equation}\label{18j31}
\begin{array}{lll}
L^2_{\{w w_* s_+,w w_*\nu_+\}}
&\stackrel{w}{\longrightarrow}& L^2_{\scp}\\
\llap{+}\Big{\uparrow}\Big\downarrow\rlap{--} & &
\llap{+}\Big{\uparrow}\Big\downarrow\rlap{--}\\
L^2_{\{w^{-1} w_*^{-1} s_-,w^{-1}
w_*^{-1}\nu_-\}}&\stackrel{w_*^{-1}}{\longrightarrow}& L^2_{\scm}
\end{array}
\end{equation}
Here the horizontal arrows are related to the unitary multiplication
operators and the vertical arrows are related to two
different $\pm$-duality mappings.
\end{lemma}

\begin{proof}
Note that both $w$ and $w_*^{-1}$ are well defined on $\bbR\cup\Lambda$.
Evidently,
$wf\in L^2_{\scp}$ means that $f\in L^2_{\{w w_* s_+,w w_*\nu_+\}}$.
Since $|w(\lambda)|=1$, $\lambda\in \bbR$, we have that 
$\{w^{-1} w_*^{-1} s_-,w^{-1}w_*^{-1}\nu_-\}$ are
minus--scattering data for $\{w w_* s_+,w w_*\nu_+\}$
if $\scm$ corresponds to $\scp$. In other words, the $s$-function remains
the  same for both sets of scattering data. Then we use definitions
\eqref{4.10d}, \eqref{5.10d}.
\end{proof}

Let $b_{\lambda_0}=\frac{\lambda-\lambda_0}{\lambda-\bar\lambda_0}$.
Note that $(b_{\lambda_0})_*=b_{-\bar\lambda_0}$.

\begin{lemma}\label{l2.1f5}
We have
\begin{equation}\label{2.1f1}
(k_{\scp}(\lambda,\lambda_0))^-=
\frac{1}{s(-\bar\lambda_0)}
\frac{b^{-1}_{-\bar\lambda_0}(\lambda)}{2\Im \lambda_0}
\frac{\hat k_{\{b^{-1}_{\lambda_0}
b^{-1}_{-\bar\lambda_0}s_-,b^{-1}_{\lambda_0}
b^{-1}_{-\bar\lambda_0}\nu_-\}}(\lambda,-\bar\lambda_0)}
{\hat k_{\{b^{-1}_{\lambda_0}
b^{-1}_{-\bar\lambda_0}s_-,b^{-1}_{\lambda_0}
b^{-1}_{-\bar\lambda_0}\nu_-\}}(-\bar\lambda_0,-\bar\lambda_0)},
\end{equation}
and, consequently,
\begin{equation}\label{2.1af1}
k_{\scp}(\lambda_0,\lambda_0)
\hat k_{\{b^{-1}_{\lambda_0}
b^{-1}_{-\bar\lambda_0}s_-,b^{-1}_{\lambda_0}
b^{-1}_{-\bar\lambda_0}\nu_-\}}(-\bar\lambda_0,-\bar\lambda_0)
=\frac{1}{s(-\bar\lambda_0)s(\lambda_0)}\frac{1}{(2\Im\lambda_0)^2}.
\end{equation}
\end{lemma}
\begin{proof} First we note that the following one-dimensional
spaces coincide
\begin{equation*}
\{(k_{\scp}(\lambda,\lambda_0)\}^-=
\{b^{-1}_{-\bar\lambda_0}
\hat k_{\{b^{-1}_{\lambda_0}
b^{-1}_{-\bar\lambda_0}s_-,b^{-1}_{\lambda_0}
b^{-1}_{-\bar\lambda_0}\nu_-\}}(\lambda,-\bar\lambda_0)\}.
\end{equation*}
This follows immediately from Theorem \ref{t1.3},
but we prefer to give a formal proof.
Starting with the orthogonal decomposition
\begin{equation*}
\{k_{\scp}(\lambda,\lambda_0)\}=
H^2_{\scp}\ominus b_{\lambda_0}H^2_{\{b_{\lambda_0}
b_{-\bar\lambda_0}s_+,b_{\lambda_0}
b_{-\bar\lambda_0}\nu_+\}}
\end{equation*}
we have
\begin{equation*}
\{k_{\scp}(\lambda,\lambda_0)\}^-=
(H^2_{\scp})^-\ominus (b_{\lambda_0}H^2_{\{b_{\lambda_0}
b_{-\bar\lambda_0}s_+,b_{\lambda_0}
b_{-\bar\lambda_0}\nu_+\}})^-,
\end{equation*}
or, due to \eqref{18j31},
\begin{equation*}
\{k_{\scp}(\lambda,\lambda_0)\}^-=
(H^2_{\scp})^-\ominus b^{-1}_{-\bar\lambda_0}(H^2_{\{b_{\lambda_0}
b_{-\bar\lambda_0}s_+,b_{\lambda_0}
b_{-\bar\lambda_0}\nu_+\}})^-.
\end{equation*}
Now we use Theorem \ref{t1.3}
\begin{equation*}
\begin{split}
\{k_{\scp}(\lambda,\lambda_0)\}^-=&
(L^2_{\scm}\ominus \hat H^2_{\scm})\\
\ominus&
b^{-1}_{-\bar\lambda_0}(L^2_{\{b^{-1}_{\lambda_0}
b^{-1}_{-\bar\lambda_0}s_-,b^{-1}_{\lambda_0}
b^{-1}_{-\bar\lambda_0}\nu_-\}}
\ominus
\hat H^2_{\{b^{-1}_{\lambda_0}
b^{-1}_{-\bar\lambda_0}s_-,b^{-1}_{\lambda_0}
b^{-1}_{-\bar\lambda_0}\nu_-\}})\\
=&b^{-1}_{-\bar\lambda_0}
(\hat H^2_{\{b^{-1}_{\lambda_0}
b^{-1}_{-\bar\lambda_0}s_-,b^{-1}_{\lambda_0}
b^{-1}_{-\bar\lambda_0}\nu_-\}}
\ominus b_{-\bar\lambda_0}\hat H^2_{\scm}).
\end{split}
\end{equation*}
Thus
\begin{equation}\label{2.2f1}
(k_{\scp}(\lambda,\lambda_0))^-=
C b^{-1}_{-\bar\lambda_0}
\hat k_{\{b^{-1}_{\lambda_0}
b^{-1}_{-\bar\lambda_0}s_-,b^{-1}_{\lambda_0}
b^{-1}_{-\bar\lambda_0}\nu_-\}}(\lambda,-\bar\lambda_0).
\end{equation}

The essential part of the lemma deals with the constant $C$. We
calculate the scalar product
\begin{equation*}
\left\langle k_{\scp}(\lambda,\lambda_0), 
\frac{iB(\lambda)}{\lambda-\bar\lambda_0}
\right\rangle_{\scp}.
\end{equation*}
On the one hand, since $\frac{iB(\lambda)}{\lambda-\bar\lambda_0}$
belongs to the intersection of $L^2_{\scp}$ with $H^2$,
we can use the reproducing property of $k_{\scp}$:
\begin{equation}\label{2.3f1}
\left\langle k_{\scp}(\lambda,\lambda_0), 
\frac{iB(\lambda)}{\lambda-\bar\lambda_0}
\right\rangle_{\scp}=
\overline{\frac{B(\lambda_0)}{2\Im\lambda_0}}=
\frac{B(-\bar\lambda_0)}{2\Im\lambda_0}.
\end{equation}
On the other hand we can reduce the given scalar product to the
scalar product in the standard $H^2$. Since $B(\lambda_k)=0$,
the
$\nu$--component disappears and we get 
\begin{equation*}
\begin{split}
\frac 1 2 &\left\langle 
\begin{bmatrix} 1&\bar s_+\\
s_+&1\end{bmatrix}(\lambda)
\begin{bmatrix}
k_{\scp}(\lambda,\lambda_0)\\ -k_{\scp}(-\bar\lambda,\lambda_0)
\end{bmatrix},
\begin{bmatrix}
\frac{iB(\lambda)}{\lambda-\bar\lambda_0}\\
\frac{iB(-\bar\lambda)}{\bar\lambda+\bar\lambda_0}
\end{bmatrix}
\right\rangle\\=&
\left\langle s(\lambda) (k_{\scp}(\lambda,\lambda_0))^-, 
\frac{i\overline{B(\lambda)}}{\lambda+\bar\lambda_0}
\right\rangle.
\end{split}
\end{equation*}
Substituting here \eqref{2.2f1} and using $s=s_e/B$ we come to
\begin{equation*}
C\left\langle s_e(\lambda)
\hat k_{\{b^{-1}_{\lambda_0}
b^{-1}_{-\bar\lambda_0}s_-,b^{-1}_{\lambda_0}
b^{-1}_{-\bar\lambda_0}\nu_-\}}(\lambda,-\bar\lambda_0), 
b_{-\bar\lambda_0}(\lambda)\frac{i}{\lambda+\bar\lambda_0}
\right\rangle.
\end{equation*}
Since $s_e(\lambda)
\hat k_{\{b^{-1}_{\lambda_0}
b^{-1}_{-\bar\lambda_0}s_-,b^{-1}_{\lambda_0}
b^{-1}_{-\bar\lambda_0}\nu_-\}}(\lambda,-\bar\lambda_0)$
belongs to $H^2$ and
$b_{-\bar\lambda_0}(\lambda)\frac{i}{\lambda+\bar\lambda_0}=
\frac{i}{\lambda+\lambda_0}$ is the reproducing kernel of $H^2$,
relation \eqref{2.3f1} yields
\begin{equation*}
C s_e(-\bar\lambda_0)
\hat k_{\{b^{-1}_{\lambda_0}
b^{-1}_{-\bar\lambda_0}s_-,b^{-1}_{\lambda_0}
b^{-1}_{-\bar\lambda_0}\nu_-\}}(-\bar\lambda_0,-\bar\lambda_0)=
\frac{B(-\bar\lambda_0)}{2\Im\lambda_0}.
\end{equation*}
Thus \eqref{2.1f1} is proved. Comparing the norms of these
vectors and taking into account that the $-$-map is an isometry
we get \eqref{2.1af1}.
\end{proof}
 
As a consequence of the above lemma, we have
$$
b_{-\bar\l_0}(e^{i\l x}K_\scpx(.,\l_0))^-\in \hat H^2_\scmb,\quad x\ge 0.\\
$$ 
Indeed, using diagram \eqref{18j31}, we get for $x\ge 0$
\begin{eqnarray*}
&&(e^{-i\l x}K_\scpxx(.,\l_0))^-_{s_-}=e^{i\l x}(K_\scpxx(.,\l_0))^-_{s_- 
e^{2i\l x}}\\
&&=C(\l_0)e^{i\l x}\hat K_\scmxb(.,-\bar\l_0)
\end{eqnarray*}
by \eqref{2.1f1}.  So, the latter function is in $\hat H^2_\scmb$. 

For discrete measures $\nu_\pm$ \eqref{6.9d}, let $\nu_{\pm N}$ be their
truncations
$$
\nu_{\pm N}=\sum^N_{k=1}\nu_\pm(\l_k)\d_{\l_k}.
$$
We say few more words about spaces $H^2_{\{s_\pm,\nu_\pm\}},\hat
H^2_{\{s_\pm,\nu_\pm\}}$ and $H^2_{\{s_\pm,\nu_{\pm N}\}},\hat
H^2_{\{s_\pm,\nu_{\pm N}\}}$.
Recall that $H^2_\scp\subset \hat H^2_\scp$. 

\begin{lemma}\label{l3}\hfill
\begin{itemize}
\item[\it i)] Let $||s_+||_\infty<1$ (or, what is the same, $\inf_\br |s_e|>0$). 
Then
$$
H^2_\scp=\hat H^2_\scp,
$$
and, consequently, $K_\scp(.,\l_0)=\hat K_\scp(.,\l_0)$.
\item[\it ii)] We always have
$$
K_\scp(\l_0,\l_0)\le \hat K_\scp(\l_0,\l_0).
$$
The equality above takes place if and only if $K_\scp(.,\l_0)=\hat 
K_\scp(.,\l_0)$.
\item[\it iii)] Obviously,
$$
H^2_\scp\subset H^2_\scpn,\quad \hat H^2_\scp\subset \hat H^2_\scpn,
$$
and
\begin{eqnarray*}
K_\scp(\l_0,\l_0)&\le& K_\scpn(\l_0,\l_0),\quad\\
\hat K_\scp(\l_0,\l_0)&\le& \hat K_\scpn(\l_0,\l_0).
\end{eqnarray*}
As before, the inequalities become equalities if and only if the
corresponding reproducing kernels coincide.
\end{itemize}
\end{lemma}

\begin{proof} To prove $i)$, we only have to show the 
inverse inclusion. 
Suppose that $f\in \hat H^2_\scp$. By Definition \ref{d11}, $f\in L^2_\scp$ and 
$s_ef\in H^2$. Since $s_e, 1/s_e\in H^\infty$, we see $f\in 
H^2$ and hence $f\in H^2_\scp$. The claim about the reproducing kernels 
is trivial.

The inequality in $ii)$ of course follows from inclusion $H^2_\scp\subset 
\hat H^2_\scp$. Consider a system $\{f_n\}_{n\in \bz_+}, f_n=b^n_{\l_0}\hat 
K_\scpbn(.,\l_0)$. This is an orthonormal basis in $\hat H^2_\scp$. We have 
$K_\scp(.,\l_0)\in \hat H^2_\scp$ and $||K_\scp$ $(.,\l_0)||_\scp=1$. So
$$
K_\scp(.,\l_0)=\sum_n a_n f_n
$$ 
and $a_0=K_\scp(\l_0,\l_0)/\hat K_\scp(\l_0,\l_0)$. Obviously, $|a_0|^2\le 1$ 
and claim $ii)$ is proved.

Let us have a look at $iii)$. The first inclusion follows from the
fact that for $f\in H^2(\bc_+)$
$$
||B_{N,\infty} f||_\scpn\le ||B_{N,\infty} f||_\scp.
$$
The second one follows from Definition \ref{d11} of $\hat H^2_\scp$. The 
inequalities for the
reproducing kernels are corollaries of these inclusions; to prove them
just argue as in $ii)$.
\end{proof}

In particular, we have
\begin{eqnarray*}
&&K_\scpxx(\l_0,\l_0)\, K_\scmxb(-\bar\l_0,-\bar\l_0)\\
&&=\frac1{|s(\l_0)|(2\im\l_0)}.
\end{eqnarray*}
under assumptions $i)$ of the above lemma.

We denote by $P_\scp$ the orthogonal projector from $L^2_\scp$ on $H^2_\scp$.  
Furthermore, $P_{x,\scp}$ and $\hat P_{x, \scp}$ are orthogonal projectors on 
$H^2_\scp(x)$ and $\hat H^2_\scp(x)$, correspondingly.

\begin{lemma}\label{l31} We have for any $f\in L^2_\scp$:
\begin{eqnarray}
i)\  \lim_{x\to -\infty} P_{x,\scp} f=f,&& \lim_{x\to -\infty} \hat P_{x,\scp} 
f=f \label{ee05}\\
ii)\  \lim_{x\to +\infty} P_{x,\scp} f=0,&& \lim_{x\to +\infty} \hat P_{x,\scp} 
f=0 \label{ee06}
\end{eqnarray}
\end{lemma}
Symbolically, we may say that
\begin{eqnarray*}
&i)\ &  \lim_{x\to -\infty} e^{i\l x} H^2_\scpx=L^2_\scp,
\lim_{x\to -\infty} e^{i\l x} \hat H^2_\scpx=L^2_\scp,\\
&ii)& \lim_{x\to +\infty} e^{i\l x} H^2_\scpx=\{0\}, 
\lim_{x\to +\infty} e^{i\l x} \hat H^2_\scpx=\{0\}.\\
\end{eqnarray*} 
\begin{proof}
We prove  the first equality in \eqref{ee06}; the argument for the second 
equality is likewise. Relations in \eqref{ee05} drop by duality, since
$$
L^2_\scp=e^{i\l x} H^2_\scpx\oplus \lt(e^{-i\l x}\hat H^2_\scmxx\rt)^+.
$$
Obviously, $e^{i\l x_2} H^2_{\{s_+e^{2i\l x_2},\, \nu_+e^{2i\l x_2}\}}\subset 
e^{i\l x_1} H^2_{\{s_+e^{2i\l x_1},\, \nu_+e^{2i\l x_1}\}}$ for $x_1\le x_2$ and 
so $k_\scp(\l_0,\l_0; x)=e^{-2\im \l_0 x}k_\scpx (\l_0,\l_0)$ is decreasing with 
respect to $x\in\br$. We have to prove that
$$
\lim_{x\to+\infty} e^{-2\im \l_0 x}k_\scpx (\l_0,\l_0)=0,
$$
which is trivial since the second factor tends to $k(\l_0,\l_0)$ by Lemma 
\ref{l6}.
\end{proof}

\begin{corollary}\label{c12}
We have
$$
\lim_{x\to-\infty} e^{-2\im\l_k x} k_\scpx(\l_k,\l_k)=\frac 1{\nu_+(\l_k)}.
$$
\end{corollary}
\begin{proof} Let us consider $g_k=(1/\nu_+(\l_k))\, \d_{\l_k}$. Recall that 
$k_\scp(.,\l_0; x)=e^{ix(\l-\bar\l_0)}$ $k_\scpx(.,\l_0)$. Hence, we obtain for 
a $f\in H^2_\scp(x)$
$$
(f,g_k)_\scp=\frac{f(\l_k)}{\nu_+(\l_k)}\, \nu_+(\l_k)=f(\l_k).
$$
On the other hand,
$$
f(\l_k)=(f, k_\scp(.,\l_k; x))_\scp=(f, P_{x, \scp} g_k)_\scp.
$$
By  Lemma \ref{l31},
$$
\lim_{x\to-\infty} ||k_\scp(.,\l_k; x)||^2_\scp=||g_k||^2_\scp
$$
which becomes the claim of the corollary if we write the norms explicitly.
\end{proof}

\subsection{Proof of Theorem \ref{t2}}

\begin{lemma}\label{l1} We have
\begin{equation*}
k_\scp(.,\l_0)=\lim_{\ep\to 0+} (\ep+I+H_\scp)^{-1} k(.,\l_0),
\end{equation*}
where $H_\scp$ is the Hankel operator coming from the metric \eqref{7.9d} and 
the limit is understood in $L^2_\scp$-sense.
\end{lemma}

The argument follows \cite{VYu}, Lemma 1.2, and is omitted.

\begin{lemma}\label{l5} Let $||s_+||_\infty<1$ and $\nu_+$ be a
  measure with a finite support. Then
\begin{equation*}
\lim_{x\to+\infty} \frac{K_\scpx(\l_0,\l_0)}{K(\l_0,\l_0)}=1.
\end{equation*}
\end{lemma}

\begin{proof}
We see that 
\begin{eqnarray*}
&&|k_\scpx(\l_0,\l_0)-k(\l_0,\l_0)|^2\\
&=&|(k_\scpx(.,\l_0)-k(.,\l_0),k_\scpx(.,\l_0))_\scpx|^2\\
&=&|\big ((I+H_\scpx)\{(I+H_\scpx)^{-1}k(.,\l_0)-k(.,\l_0)\},\\
&&(I+H_\scpx)^{-1}k(.,\l_0)\big)|^2\\
&=&|(H_\scpx k(.,\l_0), (I+H_\scpx)^{-1}k(.,\l_0))|^2\\
&\le&C\Big(|(H_{s_+ e^{2i\l x}} k(.,\l_0), (I+H_\scpx)^{-1}k(.,\l_0))|^2\\
&+&|(H_{\nu_+ e^{2i\l x}} k(.,\l_0), (I+H_\scpx)^{-1}k(.,\l_0))|^2\Big)
\end{eqnarray*}
The bound for the first term is easy
$$
|\ldots|\le\frac 1{1-||s_+||_\infty} ||P_+(s_+e^{-2i\bar\l 
x}k(-\bar\l,\l_0))||_2 ||k(.,\l_0)||_2\to 0
$$
as $x\to+\infty$ by the $L^2$-Fourier theorem. Since
$F=(I+H_\scpx)^{-1}k(.,\l_0)\in H^2$ satisfies $||F||_2\le C$, we get
$|F(\l_k)|\le C/\sqrt{\im\l_k}$ and
\begin{eqnarray*}
|(H_{\nu_+ e^{2i\l x}} k(.,\l_0), (I&+&H_\scpx)^{-1}k(.,\l_0))|\\
&\le&C\sum_{k=1}^N \nu_+(\l_k) e^{-2\im\l_k 
x}\frac{|k(\l_k,\l_0)|}{\sqrt{\im\l_k}}.
\end{eqnarray*}
The right-hand side of the inequality goes to 0 as $x\to+\infty$.
\end{proof}

The following lemma is the main key to the proof of the asymptotics.
 
\begin{lemma}\label{l6} We have
\begin{eqnarray}
&&\lim_{x\to+\infty} \frac{K_\scpx(\l_0,\l_0)}{K(\l_0,\l_0)}=1, \label{ee1}\\
&&\lim_{x\to+\infty} \frac{\hat K_\scpx(\l_0,\l_0)}{K(\l_0,\l_0)}=1.\label{ee2}
\end{eqnarray}
\end{lemma}

\begin{proof}
We start with the proof of the first equality. Taking the square root of both 
sides of \eqref{2.1af1}, we see
\begin{eqnarray}
&&K_\scpx(\l_0,\l_0)\nonumber\\
&=&\frac1{2\im\l_0 |s(\l_0)|} \frac 1{\hat
  K_\scmxxb(-\bar\l_0,-\bar\l_0)}\nonumber\\
&\ge&\frac 1{2\im\l_0 |s(\l_0)|} \frac 1{\hat
  K_\scmxxbn(-\bar\l_0,-\bar\l_0)}\nonumber\\
&=&\frac{|B(\l_0)|}{|B_{1,N}(\l_0)|} K_\scpxn (\l_0,\l_0).\label{ee20}
\end{eqnarray}
Then we continue  as
\begin{eqnarray}
&&K_\scpx(\l_0,\l_0)\nonumber\\
&\ge&|B_{N,\infty}(\l_0)|(k_\scpxn(\l_0,\l_0))^{1/2} \label{ee21}\\
&\ge&|B_{N,\infty}(\l_0)|((\ep+I+H_\scpxn)^{-1}k(.,\l_0))^{1/2}(\l_0) 
\nonumber\\
&=&\frac 1{\sqrt{1+\ep}}|B_{N,\infty}(\l_0)|
K_{\{\frac{s_+}{1+\ep}e^{2i\l x},\frac{\nu_{+ N}}{1+\ep}e^{2i\l x} 
\}}(\l_0,\l_0).
\nonumber
\end{eqnarray}
Let $s_N=s_e/B_{1,N}$. We have
\begin{eqnarray}
&&K_\scpx(\l_0,\l_0)\le K_\scpxn (\l_0,\l_0) \label{ee22}\\
&=&\frac 1{2\im\l_0 |s_N(\l_0)|}\, \frac 1{\hat 
K_\scmxxbn(-\bar\l_0,-\bar\l_0)} \nonumber\\
&\le&\frac 1{2\im\l_0 |s_N(\l_0)|}\, \frac 1{K_\scmxxbn(-\bar\l_0,-\bar\l_0)}
\nonumber
\end{eqnarray}
by $ii)$, Lemma \ref{l3}. We set $s_{-,\ep}=s_-/(1+\ep), 
\nu_{-N,\ep}=\nu_{- N}/(1+\ep)$; 
the functions $s_{N,\ep}, s_{+,\ep}$ are defined by unitarity of the scattering 
matrix, and $\nu_{+N,\ep}$ is defined by $\nu_{-N,\ep}$ through relations 
\eqref{3.10d}. Notice that the support of $\nu_{+N,\ep}$ is the same as
the support of $\nu_{+ N}$ (and equals $\{\l_k\}_{k=1,N}$). Since $K$- and $\hat 
K$-kernels are the same for pairs
$\{s_{+,\ep}, \nu_{+N,\ep}\}$ and  $\{s_{-,\ep}, \nu_{-N,\ep}\}$ by
$i)$, Lemma \ref{l3}, we continue as
\begin{eqnarray*}
(\ldots)&\le&\frac{\sqrt{1+\ep}}{2\im\l_0 |s_N(\l_0)|}\frac 
1{K_\scmxbne(-\bar\l_0,-\bar\l_0)}\\
&=&\frac{\sqrt{1+\ep}\, |s_{N,\ep}(\l_0)|}{|s_N(\l_0)|}K_\scpxne(\l_0,\l_0).
\end{eqnarray*}
That is,
\begin{eqnarray*}
&&\frac{|B_{N,\infty}(\l_0)|}{\sqrt{1+\ep}}K_{\{\frac{s_+}{1+\ep}e^{2i\l
 x}, \frac{\nu_{+ N}}{1+\ep}e^{2i\l 
x} \}}(\l_0,\l_0)\le
K_\scpx(\l_0,\l_0)\\
&\le&\frac{\sqrt{1+\ep}\, |s_{N,\ep}(\l_0)|}{|s_N(\l_0)|}K_\scpxne(\l_0,\l_0).
\end{eqnarray*}
The quantities $K_{\{\frac{s_+}{1+\ep}e^{2i\l x},\frac{\nu_{+ N}}{1+\ep}e^{2i\l 
x} 
\}}(\l_0,\l_0)$ and $K_\scpxne(\l_0,\l_0)$ tend to $K(\l_0,\l_0)$ as 
$x\to+\infty$ by Lemma \ref{l5}. Remaining factors in the left- and 
right-hand side parts of the inequality go to $1$ with $\ep\to
+0, N\to+\infty$. Hence, for any $\ep'>0$ we can choose appropriate $\ep, N$ to 
have
\begin{eqnarray*}
1-\ep'&\le&\liminf_{x\to+\infty}\frac{K_\scpx(\l_0,\l_0)}{K(\l_0,\l_0)}\\
&\le&\limsup_{x\to+\infty}\frac{K_\scpx(\l_0,\l_0)}{K(\l_0,\l_0)}\le 1+\ep',
\end{eqnarray*}
and \eqref{ee1} is proved.

The proof of \eqref{ee2} is almost identical. First of all, to keep the notation 
we used to, we prove
\begin{equation}\label{ee3}
\lim_{x\to+\infty} \frac{\hat 
K_\scmx(-\bar\l_0,-\bar\l_0)}{K(-\bar\l_0,-\bar\l_0)}=1.
\end{equation}
instead of \eqref{ee2}. This is obviuosly the same thing up to changes 
$-\bar\l_0\mapsto \l_0$ and $s_-\mapsto s_+$. The 
second modification is that we estimate the value of a $\hat K$-kernel by the 
values of $K$-kernels (and not vice versa as we have just done to prove 
\eqref{ee1}).

So, as in \eqref{ee21}, we have
\begin{eqnarray*}
\hat K_\scmx(-\bar\l_0,-\bar\l_0)
&\ge& K_\scmx(-\bar\l_0,-\bar\l_0)\\
&\ge&\frac 1{\sqrt{1+\ep}} 
K_{\{\frac{s_-}{1+\ep} e^{2i\l x},\, \frac{\nu_-}{1+\ep} e^{2i\l x}\}}
(-\bar\l_0,-\bar\l_0)\\
&=&\frac{|B_{N,\infty}(\l_0)|}{\sqrt{1+\ep}} 
K_{\{\frac{s_-}{1+\ep} e^{2i\l x},\, \frac{\nu_{- N}}{1+\ep} 
e^{2i\l x}\}}(-\bar\l_0,-\bar\l_0). 
\end{eqnarray*}
The first inequality in the above estimate is $ii)$, 
Lemma \ref{l3} and the last one repeats computation
\eqref{ee20}. Similarly to 
\eqref{ee22}, we get
\begin{eqnarray*}
&&\hat K_\scmx(-\bar\l_0,-\bar\l_0)\le\hat K_\scmxn(-\bar\l_0,-\bar\l_0)\\
&=&\frac 1{2\im \l_0|s_N(\l_0)|}\frac 1{K_\scpxxbn(\l_0,\l_0)}\\
&\le&\frac{\sqrt{1+\ep}}{2\im \l_0|s_N(\l_0)|}\,
\frac 1{K_{\{\frac{s_+}{1+\ep}e^{-2i\l x}(b_{\l_0}b_{-\bar\l_0})^{-1},\, 
\frac{\nu_{+ N}}{1+\ep}e^{-2i\l x}(b_{\l_0}b_{-\bar\l_0})^{-1}\}}(\l_0,\l_0)}\\
&=&\frac{\sqrt{1+\ep}\, |s_{N,\ep}(\l_0)|}{|s_N(\l_0)|} 
\hat K_{\{s_{-,\ep} e^{2i\l x},\, \nu_{-N,\ep} e^{2i\l 
x}\}}(-\bar\l_0,-\bar\l_0)\\
&=&\frac{\sqrt{1+\ep}\, |s_{N,\ep}(\l_0)|}{|s_N(\l_0)|}
K_{\{s_{-,\ep} e^{2i\l x},\, \nu_{-N,\ep} e^{2i\l x}\}}(-\bar\l_0,-\bar\l_0)
\end{eqnarray*}
Above, the pair $\{s_{-,\ep},\nu_{-N,\ep}\}$ comes from 
$\{\frac{s_+}{1+\ep},\frac{\nu_{+ N}}{1+\ep}\}$ as explained after \eqref{ee22}.
Hence,
\begin{eqnarray*}
&&\frac{|B_{N,\infty}(\l_0)|}{\sqrt{1+\ep}} 
K_{\{\frac{s_{-}}{1+\ep} e^{2i\l x},\, \frac{\nu_{- N}}{1+\ep} e^{2i\l x}\}}
(-\bar\l_0,-\bar\l_0)
\le\hat K_\scmx(-\bar\l_0,-\bar\l_0)\\
&\le&\frac{\sqrt{1+\ep}\, |s_{N,\ep}(\l_0)|}{|s_N(\l_0)|} 
K_{\{s_{-,\ep} e^{2i\l x},\, \nu_{-N,\ep} e^{2i\l x}\}}(-\bar\l_0,-\bar\l_0). 
\end{eqnarray*}
Repeating the argument from the first part of the proof, we see that
for every $\ep'>0$
\begin{eqnarray*}
1-\ep'&\le&\liminf_{x\to+\infty}\frac{\hat 
K_\scmx(-\bar\l_0,-\bar\l_0)}{K(-\bar\l_0,-\bar\l_0)}\\
&\le&\limsup_{x\to+\infty}\frac{\hat 
K_\scmx(-\bar\l_0,-\bar\l_0)}{K(-\bar\l_0,-\bar\l_0)}\le 1+\ep',
\end{eqnarray*}
and relation \eqref{ee3} is proved.
\end{proof}

\noindent
{\it Proof of Theorem \ref{t2}.}  At present, the claim of the theorem is an 
easy consequence of 
Lemma \ref{l6}.  For an arbitrary $N$, we have 
\begin{eqnarray*}
||e^{i\l x}K_\scpx(.,\l_0)&-&e^{i\l x}\ti K(.,\l_0)||_\scp\\
&\le&||e^{i\l x}K_\scpx(.,\l_0)-B_{N,\infty} e^{i\l x}K(.,\l_0)||_\scp\\
&+&||e^{i\l x}\ti K(.,\l_0)-B_{N,\infty} e^{i\l x}K(.,\l_0)||_\scp.
\end{eqnarray*}
The claim will then follow if we prove
\begin{eqnarray}
\limsup_{x\to+\infty}&&||e^{i\l x}\ti K(.,\l_0)-B_{N,\infty} e^{i\l 
x}K(.,\l_0)||_\scp \label{ee31}\\
&\le& C_1|1-B_{N,\infty}(\l_0)| \nonumber\\
\limsup_{x\to+\infty}&&||e^{i\l x}K_\scpx(.,\l_0)-B_{N,\infty} e^{i\l 
x}K(.,\l_0)||_\scp\label{ee32}\\
&\le& C_2|1-B_{N,\infty}(\l_0)|
\nonumber
\end{eqnarray}
with some constants $C_1, C_2$.
The computation for \eqref{ee31} is easy and elementary
\begin{eqnarray*}
||e^{i\l x}\ti K(.,\l_0)&-&B_{N,\infty} e^{i\l x}K(.,\l_0)||^2_\scp\\
&\le&||e^{i\l x} (1-B_{N,\infty}) K(.,\l_0)||^2_{s_+}+||e^{i\l x} B_{N,\infty} 
K(.,\l_0)||^2_{\nu_+}
\end{eqnarray*}
The second term above obviously goes to $0$ as $x\to+\infty$; for the first one 
we have
$$
 ||e^{i\l x} (1-B_{N,\infty}) K(.,\l_0)||^2_{s_+}\le 2||e^{i\l x} 
(1-B_{N,\infty}) K(.,\l_0)||^2_2\le 4|1-B_{N,\infty}(\l_0)|.
$$ 
We pass to \eqref{ee32} now. Once again, for an arbitrary $N$,
\begin{eqnarray*}
&&||e^{i\l x}K_\scpx(.,\l_0)-B_{N,\infty} e^{i\l x}K(.,\l_0)||^2_\scp\\
&\le& ||e^{i\l x}K_\scpx(.,\l_0)||^2_\scp\\
&-&2\re (K_\scpx(.,\l_0), B_{N,\infty}K(.,\l_0))_\scpx
+||B_{N,\infty}K(.,\l_0)||^2_\scpx.
\end{eqnarray*}
By Lemma \ref{l6}, we get for the second term 
$$
\re(\ldots)=\frac{B_{N,\infty}(\l_0)K(\l_0,\l_0)}{K_\scpx(\l_0,\l_0)}\to 
B_{N,\infty}(\l_0)
$$
as $x\to+\infty $. The third term is
\begin{eqnarray*}
(\ldots)&=&||B_{N,\infty}K(.,\l_0)||^2_{s_+e ^{2i\l 
x}}+||B_{N,\infty}K(.,\l_0)||^2_{\nu_+ e^{2i\l x}}\\
&\le&||K(.,\l_0)||^2_2+||P_+ [s_+ e^{2i\l 
x}B_{N,\infty}K(.,\l_0)](-\bar\l)||_2\, ||B_{N,\infty}K(.,\l_0)||_2\\
&+&||B_{N,\infty}K(.,\l_0)||^2_{\nu_+ e^{2i\l x}}\to 1,
\end{eqnarray*}
since $||K(.,\l_0)||^2_2=1$ and the rest tends to 0 with $x\to+\infty$ (for the 
second term, this is Fourier $L^2$-theorem). So, summing up
$$
\limsup_{x\to+\infty}||e^{i\l x}K_\scpx(.,\l_0)-B_{N,\infty} e^{i\l 
x}K(.,\l_0)||_\scp\le 2\re(1-B_{N,\infty}(\l_0)),
$$
and \eqref{ee32} is proved.

The proof of \eqref{ee4} is likewise, we just have to use \eqref{ee2} instead of 
\eqref{ee1}.
\hfill $\Box$

\section{Unitary node, I}
\label{s3}
Consider the multiplication operator by $\bar v,\ 
v=\frac{\lambda^2-\lambda_0^2}{\lambda^2-\bar\lambda_0^2}$,
acting in
\begin{equation}\label{2.16d}
L^2_{\scp}=(\hat H^2_{\scm})^+\oplus
 H^2_{\scp}.
\end{equation}
\begin{lemma}\label{l2.2f5}
The multiplication operator by $\bar v$ acts 
as a unitary operator from
\begin{equation}\label{5.16d}
\{\hat k_{\scm}^+(\lambda,\lambda_0)\}\oplus
H^2_{\scp}(x)
\end{equation}
to
\begin{equation}\label{6.16d}
\{\hat k_{\scm}^+(\lambda,-\bar\lambda_0)\}\oplus
H^2_{\scp}(x).
\end{equation}
\end{lemma}
\begin{proof}
It is obvious that
the multiplication by
$\bar v=\frac{b_{-\bar\lambda_0}}
{b_{\lambda_0}}$
acts from
\begin{equation*}
\{f\in \hat H^2_{\scm}: f(\lambda_0)=0\}=
b_{\lambda_0}\hat H^2_{
\{b_{\lambda_0}b_{-\bar\lambda_0}s_-,
b_{\lambda_0}b_{-\bar\lambda_0}\nu_-
\}}
\end{equation*}
to
\begin{equation*}
\{f\in \hat H^2_{\scm}: f(-\bar\lambda_0)=0\}
=
b_{-\bar\lambda_0}\hat H^2_{
\{b_{\lambda_0}b_{-\bar\lambda_0}s_-,
b_{\lambda_0}b_{-\bar\lambda_0}\nu_-
\}}.
\end{equation*}
Therefore it acts in their orthogonal
complements \eqref{5.16d}, \eqref{6.16d}.
\end{proof}

We now recall the definition of the characteristic function
of a unitary node and its functional model. An extensive discussion of the 
subject and its application to interpolation problems can be found in 
\cite{kkhyu, kh, khyu}.

Let $K,E_1,E_2$ be Hilbert spaces and
$U$ be a unitary operator acting
from $K\oplus E_1$ to $K\oplus E_2$.
We assume that   
 $E_1$ and $E_2$ are finite-dimensional
 ($\dim E_1=\dim E_2=1$ in this section, and  $\dim E_1=\dim E_2=2$ in Section 
4).
 The characteristic function is defined by
\begin{equation}\label{7.16d}
\Theta(\zeta):=P_{E_2}U(I_{K\oplus E_1}-\zeta P_K U)^{-1}|E_1.
\end{equation}
It is a holomorphic in the unit disk $\{\zeta: |\zeta|<1\}$ contractive-valued
operator function. We make a specific assumption
that $\Theta(\zeta)$ has an analytic continuation
in the exterior of the unite disk through a certain
arc $(a,b)\subset\bbT$ by the symmetry principle
$$
\Theta(\zeta)=\Theta^*\left(\frac 1{\bar\zeta}\right)^{-1}.
$$

For $f\in K$ define
\begin{equation}\label{2.5f4}
F(\zeta):=P_{E_2}U(I-\zeta P_K U)^{-1}f.
\end{equation}
This $E_2$-valued holomorphic function belongs
to the functional space $K_{\Theta}$ with the following properties.

\begin{itemize}
\item $F(\zeta)\in H^2(E_2)$ and it has analytic continuation
through the arc $(a,b)$.
\item $F_*(\zeta):=\Theta^*(\zeta)F\left(\frac 1{\bar\zeta}\right)
\in H^2_-(E_1)$.
\item For almost every $\zeta\in \bbT$ the vector 
$\begin{bmatrix}F_*\\ F\end{bmatrix}(\zeta)$
belongs
to the image of the operator
$\begin{bmatrix} I&\Theta^*\\ \Theta & I\end{bmatrix}(\zeta)$,
and therefore the scalar product
$$
\left\langle
\begin{bmatrix} I&\Theta^*\\ \Theta &
I\end{bmatrix}^{[-1]}\begin{bmatrix}F_*\\ F\end{bmatrix},
\begin{bmatrix}F_*\\ F\end{bmatrix}
\right\rangle_{E_1\oplus E_2}
$$
is well-defined and does not depend of the choice of a preimage
(the first term in the above scalar product). Moreover,
\begin{equation}\label{2.6f4}
\int_{\bbT}
\left\langle
\begin{bmatrix} I&\Theta^*\\ \Theta &
I\end{bmatrix}^{[-1]}\begin{bmatrix}F_*\\ F\end{bmatrix},
\begin{bmatrix}F_*\\ F\end{bmatrix}
\right\rangle_{E_1\oplus E_2} dm<\infty.
\end{equation}
\end{itemize}
The integral in \eqref{2.6f4} represents the square
of the norm of $F$ in $K_{\Theta}$.

Note that  $P_K U\vert K$ becomes a certain
``standard" operator in the model space
\begin{equation}\label{2.11f23}
f\mapsto F(\zeta)\quad\Longrightarrow\quad
P_K Uf\mapsto\frac{F(\zeta)-F(0)}{\zeta},
\end{equation}
see \eqref{2.5f4}.

The following simple identity is a convenient
tool
in the forthcoming calculation.
\begin{lemma} For a unitary operator $U:K\oplus E_1\to
K\oplus E_2$
\begin{equation}\label{1.17d}
U^*P_{E_2}U(I-\zeta P_K U)^{-1}=
I+(\zeta-U^*)P_{K}U(I-\zeta P_K U)^{-1}.
\end{equation}
\end{lemma}
\begin{proof} Since $I_{K\oplus E_2}=P_K+P_{E_2}$
and $U$ is unitary we have
\begin{equation*}
U^*P_{E_2}U=
(I-\zeta P_K U)+(\zeta-U^*)P_{K}U.
\end{equation*}
Then we multiply this identity by $(I-\zeta P_K U)^{-1}$.
\end{proof}

\begin{theorem}
Let $e_1$, $e_2$ be the normalized vectors in the one-dimensional
spaces \eqref{5.16d} and \eqref{6.16d}, 
\begin{equation}\label{1.20d}
e_1(\lambda)=\frac{\hat
k_{\scm}^+(\lambda,\lambda_0)}{\sqrt{\hat
k_{\scm}(\lambda_0,\lambda_0)}},\ 
e_2(\lambda)=\frac{
\hat k^+_{\scm}(\lambda,-\bar\lambda_0)}{
\sqrt{\hat k_{\scm}(\lambda_0,\lambda_0)}}.
\end{equation}
Then the reproducing kernel
of $H^2_{\scp}$ is of the form
\begin{equation}\label{2.13f6}
k_{\scp}(\lambda,\mu)=
\frac{(v e_2)(\lambda)\overline{(v e_2)(\mu)}
-e_1(\lambda)\overline{e_1(\mu)}}{1-
v (\lambda)
\overline{v(\mu)}}.
\end{equation}
\end{theorem}

\begin{proof} First,
we are going to find the characteristic function
of the multiplication operator by $\bar v$ with respect
to decompositions \eqref{5.16d} and \eqref{6.16d}
and the corresponding functional representation
of this node.

By 
\eqref{1.20d}  
we fixed ``basises'' in the one-dimensional
spaces. 
So, instead of the operator we get a scalar function
$\theta(\zeta)$:
\begin{equation}\label{3.20d}
\Theta(\zeta) e_1:=
P_{E_2}U
(I-\zeta P_K U)^{-1}e_1
=e_2\theta(\zeta).
\end{equation}
We substitute \eqref{3.20d} in \eqref{1.17d}
\begin{equation}\label{2.14f5}
v(\lambda)e_2(\lambda)
\theta(\zeta)
=
e_1(\lambda)
+(\zeta-v(\lambda))(P_{K}U(I-\zeta P_K U)^{-1}e_1)
(\lambda).
\end{equation}
Recall an important property of
$\hat k_{\scm}^+(\lambda,\lambda_0)$: it has analytic continuation in
the upper half-plane with the only pole at $-\bar\lambda_0$
(see Lemma \ref{l2.1f5}).
Therefore all terms in \eqref{2.14f5} are analytic in $\lambda$ and
we can choose $\lambda$ satisfying $v(\lambda)=\zeta$. Then we obtain
the characteristic function in terms of
the reproducing kernels
\begin{equation}\label{2.16f8}
\theta(v(\lambda))
=\frac
{e_1(\lambda)}{v(\lambda)e_2(\lambda)}.
\end{equation}

Similarly for $f\in K=H^2_{\scp}$ we define the scalar function
$F(\zeta)$ by
\begin{equation}\label{2.17f8}
P_{E_2}U(I-\zeta P_{K} U)^{-1}f
=e_2 F(\zeta).
\end{equation}
Using again \eqref{1.17d} we get
\begin{equation*}
v(\lambda)e_2(\lambda)
F(\zeta)
=
f(\lambda)
+(\zeta-v(\lambda))(P_{K}U(I-\zeta P_K U)^{-1}f)
(\lambda).
\end{equation*}
Therefore,
\begin{equation}\label{2.18f8}
F(v(\lambda))
=\frac
{f(\lambda)}{v(\lambda)e_2(\lambda)}.
\end{equation}

Now we are in a position to get \eqref{2.13f6}.
Indeed, by \eqref{2.17f8} and \eqref{2.18f8} we proved that the vector
$$
P_{K}(I-\overline{v(\mu)} U^*P_K)^{-1}U^* e_2
\overline{v(\mu)e_2(\mu)}
$$
is the reproducing kernel of $K=H^2_{\scp}$ with respect
to $\mu$, $|v(\mu)|<1$.
Using the Darboux identity 
\begin{equation*}
P_{E_2}U(I-\zeta P_{K} U)^{-1}
P_{K}(I-{\bar\zeta_0} U^*P_K)^{-1}U^*| E_2=
\frac{I-\Theta(z)\Theta^*(\zeta_0)}{1-\zeta\bar\zeta_0}
\end{equation*}
(in this setting this is a simple and pleasant  exercise)
we obtain
\begin{equation*}
k_{\scp}(\lambda,\mu)=
v(\lambda)e_2(\lambda)
\frac{I-\theta(v(\lambda))\overline{\theta(v(\mu))}}
{1-v(\lambda)\overline{v(\mu)}}
\overline{v(\mu)e_2(\mu)}
\end{equation*}
for $|v(\lambda)|<1$, $|v(\mu)|<1$. By analyticity and
\eqref{2.16f8} we have that relation \eqref{2.13f6}  holds
for all $\l,\mu\in \bc_+$. 
\end{proof}

\begin{corollary} 
The following Wronskian-type identity is satisfied
for the reproducing kernels
\begin{equation}\label{wif2}
\left|
\begin{matrix}
(s e_2^{-})(\mu)&
 (s e_1^{-})(\mu)\\
e_2(\mu)&
e_1(\mu)
\end{matrix}\right |=
\frac 1 i(\log v(\mu))',\ \Im\mu>0.
\end{equation}
\end{corollary}

\begin{proof}
To be brief, we write $k^-_{\scp}(.,.)$ instead of $(k^-_{\scp}(.,.))^-$. So we 
multiply $k^-_{\scp}(\lambda,-\bar\mu)$ by $b_{\mu}(\lambda)$
and calculate the resulting function of $\lambda$ at
$\lambda=\mu$. By \eqref{2.1f1} we get
\begin{equation}\label{2.15f3}
\{b_{\mu}(\lambda)k^-_{\scp}(\lambda,-\bar\mu)\}_{\lambda=\mu}
=\frac{1}{s(\mu)2\Im\mu }.
\end{equation}
Now we make the same calculation  using  representation
\eqref{2.13f6}. Since
\begin{equation*}
k^-_{\scp}(\lambda,-\bar\mu)\\
=\frac
{-v(\mu)}
{v(\lambda)-v(\mu)}\left|
\begin{matrix}
v(\lambda) e_2^{-}(\lambda)&
 e_1^{-}(\lambda)\\
\overline{e_1(-\bar\mu)}&
\overline{v(-\bar\mu)e_2(-\bar\mu)}
\end{matrix}\right |,
\end{equation*}
we get in combination with \eqref{2.15f3}
\begin{equation*}
 -i\frac{v'(\mu)}
{v(\mu)s(\mu)}
= 
\left|
\begin{matrix}
v(\mu) e_2^{-}(\mu)&
 e_1^{-}(\mu)\\
\overline{e_1(-\bar\mu)}&
v^{-1}(\mu)\overline{e_2(-\bar\mu)}
\end{matrix}\right |.
\end{equation*}
By the symmetry $\overline{\hat k_{\scm}(\lambda,\lambda_0)}
=\hat k_{\scm}(-\bar\lambda,-\bar\lambda_0)$, we have
$\overline{e_2(-\bar\mu)}=e_1(\mu)$. Thus
\eqref{wif2} is proved.
\end{proof}

\begin{corollary} 
Let $\mu\in \bbR_+$  and as before $\Re \lambda_0>0$, then
\begin{equation}\label{wif8}
|e_2(\mu)|^2-
 |e_1(\mu)|^2=
\frac 1 i(\log v(\mu))'.
\end{equation}
\end{corollary}
\begin{proof} All terms in \eqref{wif2} have boundary values.
Recall that on the real axis
$(s e_{1,2}^-)(\mu)=(s_-e_{1,2})(\mu)-e_{1,2}(-\bar\mu)$. Then
use again the symmetry of the reproducing kernel.
\end{proof}

We finish this section with a translation of the relation
$$
\Vert f\Vert^2_{\scp}
=\left\Vert \frac{f}{ve_2}\right\Vert^2_{K_{\theta}}
$$
(\eqref{2.18f8} is a unitary map from $H^2_{\scp}$ to $K_{\theta}$)
to the following proposition.

\begin{theorem} Let 
\begin{equation}
s_+^{\theta}(\lambda):=\frac{e_2(-\bar\lambda)}{e_2(\lambda)},
\quad \lambda\in \bbR_+,
\end{equation}
extended by the symmetry $s_+^{\theta}(-\bar\lambda)
=\overline{s_+^{\theta}(\lambda)}$ to the whole $\br$. Let
$\nu_+^{\theta}$ be a positive measure on the imaginary half-axis
\begin{equation}
d\nu _+^{\theta}(\lambda):=\frac{|dv(\lambda)|}{2\pi |e_2(\lambda)|^2},
\quad \lambda\in i\bbR_+.
\end{equation}
Then
\begin{equation}\label{2.24f9}
\begin{split}
||f||^2_{\scp}=& \int_{i\bbR_+}|f(\lambda)|^2
d\nu^{\theta}_+(\lambda)\\
               +&\frac 1 {4\pi}
               \int_{\bbR}\begin{bmatrix}
               \overline{f(\lambda)}& 
               \overline{-f(-\bar\lambda)}
               \end{bmatrix}
               \begin{bmatrix}
               1&  
               \overline{s^{\theta}_+(\lambda)}\\
               s^{\theta}_+(\lambda)& 1
               \end{bmatrix}
               \begin{bmatrix}
               {f(\lambda)}\\
               {-f(-\bar\lambda)}
               \end{bmatrix}d\lambda
               \end{split}
\end{equation}
for all $f\in H^2_{\scp}$. In other words
\begin{equation*}
id: H^2_{\scp}\to H^2_{\{s^{\theta}_+,\nu^{\theta}_+\}}
\end{equation*}
is an isometry.
\end{theorem}

\begin{proof} We use definition of the scalar product
in $K_{\theta}$, relations \eqref{2.16f8}, \eqref{2.18f8},
and \eqref{wif8}.
\end{proof}

\section{Unitary node, II: a canonical system}
\label{s4}
In this section we associate a canonical system
(see \cite{dB, Rem}) with the given chain 
$\{H^2_{\scp}(x)\}_{x\in\bbR}$ of subspaces of $L^2_{\scp}$ . 

\subsection{Characteristic function of a unitary node
and transfer matrix. Definitions}
This time we consider the unitary multiplication operator by $\bar v$,
$v=\frac{\lambda^2-\lambda_0^2}{\lambda^2-\bar\lambda_0^2}$,
with respect to the decomposition
\begin{equation}\label{2.16d1}
L^2_{\scp}=(\hat H^2_{\scm})^+\oplus K_{\scp}(x)\oplus
 H^2_{\scp}(x).
\end{equation}
Actually this is definition of the space  $K_{\scp}(x)$.

The following lemma is similar to Lemma \ref{l2.2f5}.
\begin{lemma}
The multiplication operator by $\bar v$ acts from
\begin{equation}\label{5.16d2}
\{\hat k_{\scm}^+(\lambda,\lambda_0)\}\oplus
K_{\scp}(x)\oplus\{k_{\scp}(\lambda,\lambda_0;x)\}
\end{equation}
to
\begin{equation}\label{6.16d2}
\{\hat k_{\scm}^+(\lambda,-\bar\lambda_0)\}\oplus
K_{\scp}(x)\oplus\{k_{\scp}(\lambda,-\bar\lambda_0;x)\}.
\end{equation}
\end{lemma}

We define normalized vectors that form orthonormal basises
in $E_1$ and $E_2$ 
\begin{equation}\label{1.20d2}
 e^{(1)}_1(\lambda)=\frac{\hat
k_{\scm}^+(\lambda,\lambda_0)}{||\hat
k_{\scm}^+(\lambda,\lambda_0)||},\ 
e^{(1)}_2(\lambda)=\frac{
k_{\scp}(\lambda,\lambda_0;x)}{||k_{\scp}(\lambda,\lambda_0;x)||};
\end{equation}
and
\begin{equation}\label{2.20d}
e^{(2)}_1(\lambda)=\frac{\hat
k_{\scm}^+(\lambda,-\bar\lambda_0)}{||\hat
k_{\scm}^+(\lambda,-\bar\lambda_0)||},\ 
e^{(2)}_2(\lambda)=\frac{
k_{\scp}(\lambda,-\bar\lambda_0;x)}{||k_{\scp}(\lambda,
-\bar\lambda_0;x)||}.
\end{equation}
We point out that the vectors $e_2^{(i)}(\lambda)$, $i=1,2$, depend
also on $x$ and $e_1^{(i)}(\lambda)$, $i=1,2$, do not.
 
 Generally for an operator 
 $A:H_1\oplus H_2\to \tilde H_1\oplus\tilde H_2$ 
 its Potapov-Ginzburg transform
 $\tilde A:H_1\oplus \tilde H_2\to \tilde H_1\oplus H_2$ is defined by \cite{po, 
ka}
 \begin{equation*}
 \begin{bmatrix} y_1\\ x_2\end{bmatrix}
 = \tilde A\begin{bmatrix} x_1\\ y_2\end{bmatrix},
\quad \text{where}\quad
 \begin{bmatrix} y_1\\ y_2\end{bmatrix}
 =  A\begin{bmatrix} x_1\\ x_2\end{bmatrix}.
 \end{equation*}
In terms of the block decomposition of 
$A=\begin{bmatrix} A_{11}&A_{12}\\ A_{21}& A_{22}\end{bmatrix}$ 
we have
\begin{equation*}
 \begin{bmatrix} A_{11}&0\\ A_{21}& -I\end{bmatrix}
 \begin{bmatrix} x_1\\ y_2\end{bmatrix}=
 \begin{bmatrix} I&-A_{12}\\ 0&-A_{22}\end{bmatrix}
 \begin{bmatrix} y_1\\ x_2\end{bmatrix}.
 \end{equation*}
Therefore,
\begin{equation}\label{4.6f24}
\tilde A=
 \begin{bmatrix} I&-A_{12}\\ 0&-A_{22}\end{bmatrix}^{-1}
 \begin{bmatrix} A_{11}&0\\ A_{21}& -I\end{bmatrix}=
 \begin{bmatrix} A_{11}- A_{12}A_{22}^{-1}A_{21}
 &A_{12}A_{22}^{-1}\\ -A_{22}^{-1}A_{21}&
A_{22}^{-1}\end{bmatrix}.
 \end{equation}
The transformation is well-defined if 
$A_{22}$ is invertible. Note, that if $A$ is unitary, 
$$
\Vert y_1\Vert^2+\Vert y_2\Vert^2=
\Vert x_1\Vert^2 +\Vert x_2\Vert^2,
$$
 then $\tilde A$ preserves the indefinite metric
 $$
\Vert y_1\Vert^2 -\Vert x_2\Vert^2
=\Vert x_1\Vert^2-\Vert y_2\Vert^2.
$$
 
For the unitary node given by the multiplication
operator by $\bar v$ and decompositions \eqref{5.16d2},
\eqref{6.16d2}:
\begin{equation}\label{4.7f24}
U:(K\oplus\{e_1^{(1)}\})\oplus
\{e^{(1)}_2\}\to
(K\oplus\{e_2^{(2)}\})\oplus
\{e^{(2)}_1\}
\end{equation}
we define the $j$-unitary node
\begin{equation}\label{4.8f24}
\tilde U:(K\oplus\{e_1^{(1)}\})\oplus
\{e^{(2)}_1\}\to
(K\oplus\{e_2^{(2)}\})\oplus
\{e^{(1)}_2\}
\end{equation}
by \eqref{4.6f24}, separating in this way $x$-depending ``channels".

The characteristic operator-valued function for the node \eqref{4.7f24}
is
\begin{equation}\label{4.9f24}
\Theta(\zeta):=P_{E_2}U(I-\zeta P_K U)^{-1}\vert E_1,
\end{equation}
and its matrix with respect to the chosen basises is
\begin{equation}\label{3.20d2}
\Theta(\zeta)
\begin{bmatrix} e^{(1)}_1(\lambda)&
e^{(1)}_2(\lambda)
\end{bmatrix}\begin{bmatrix}c_1\\c_2\end{bmatrix}
=
\begin{bmatrix} e^{(2)}_2(\lambda)&
e^{(2)}_1(\lambda)
\end{bmatrix}
\theta(\zeta)
\begin{bmatrix}c_1\\c_2\end{bmatrix},
\end{equation}
where
$$
\theta(\zeta)=
\begin{bmatrix}\theta_{11}&\theta_{12}\\
\theta_{21}&\theta_{22}\end{bmatrix}(\zeta).
$$
Respectively, its functional representation is of the form
\begin{equation}\label{4.7f18}
P_{E_2}U(I-\zeta P_K U)^{-1}f
=
\begin{bmatrix} e^{(2)}_2(\lambda)&
e^{(2)}_1(\lambda)
\end{bmatrix}
\begin{bmatrix}F_1\\F_2\end{bmatrix}(\zeta),
\end{equation}
for $f\in K_{\scp}(x)$.

The transfer matrix is actually the characteristic matrix function of
the node \eqref{4.8f24}. Having \eqref{4.9f24}, \eqref{3.20d2},
we rewrite \eqref{4.7f24} in the block form as
\begin{equation}
U\begin{bmatrix}\zeta k(\zeta)\\ \begin{bmatrix}
c_1\\c_2\end{bmatrix}\end{bmatrix}=
\begin{bmatrix} k(\zeta)\\ 
\begin{bmatrix}\theta_{11}&\theta_{12}\\
\theta_{21}&\theta_{22}\end{bmatrix}(\zeta)
\begin{bmatrix}
c_1\\c_2\end{bmatrix}\end{bmatrix}.
\end{equation}
Consequently, we get for $\tilde U$:
\begin{equation}
\tilde U\begin{bmatrix}\zeta k(\zeta)\\ 
\begin{bmatrix}
1&0\\ \theta_{21}(\zeta)&\theta_{22}(\zeta)
\end{bmatrix}
\begin{bmatrix}
c_1\\c_2\end{bmatrix}\end{bmatrix}=
\begin{bmatrix} k(\zeta)\\
\begin{bmatrix}
 \theta_{11}(\zeta)&\theta_{12}(\zeta)\\ 0&1
\end{bmatrix}
\begin{bmatrix}
c_1\\c_2\end{bmatrix}\end{bmatrix}.
\end{equation}
Therefore the transfer matrix $\fA(\zeta)$ of the $j$-node is related to
$\theta(\zeta)$ by
$$
\fA(\zeta)\begin{bmatrix}
1&0\\ \theta_{21}(\zeta)&\theta_{22}(\zeta)
\end{bmatrix}=\begin{bmatrix}
 \theta_{11}(\zeta)&\theta_{12}(\zeta)\\ 0&1
\end{bmatrix}.
$$
Thus
\begin{equation}\label{4.14f24}
\fA(\zeta)=\begin{bmatrix}
 \theta_{11}(\zeta)&\theta_{12}(\zeta)\\ 0&1
\end{bmatrix}
\begin{bmatrix}
1&0\\ \theta_{21}(\zeta)&\theta_{22}(\zeta)
\end{bmatrix}^{-1}=
\begin{bmatrix} 1&
 -\theta_{12}(\zeta)\\ 0&-\theta_{22}(\zeta)
\end{bmatrix}^{-1}
\begin{bmatrix}
 \theta_{11}(\zeta)&0\\ \theta_{21}(\zeta)&-1
\end{bmatrix}.
\end{equation}

\subsection{Calculating $\theta$ and $\fA$}
We are following the same lines as in Section \ref{s3}.
Let us substitute \eqref{3.20d2} into \eqref{1.17d}
\begin{equation*}
\begin{split}
v(\lambda)\begin{bmatrix} e^{(2)}_2(\lambda)&
e^{(2)}_1(\lambda)
\end{bmatrix}
\theta(\zeta)
=&
\begin{bmatrix} e^{(1)}_1(\lambda)&
e^{(1)}_2(\lambda)
\end{bmatrix}\\
+&(\zeta-v(\lambda))
\left(P_{K}U(I-\zeta P_K U)^{-1}
\begin{bmatrix} e^{(1)}_1&
e^{(1)}_2
\end{bmatrix}\right)(\lambda).
\end{split}
\end{equation*}
All terms here are analytic in $\lambda$ and
we can choose $\lambda\in\bc_+$ with the property $v(\lambda)=\zeta$. Then we 
get
\begin{equation}\label{4.20d}
v(\lambda)
\begin{bmatrix} e^{(2)}_2(\lambda)&
e^{(2)}_1(\lambda)
\end{bmatrix}
\theta(v(\lambda))
=
\begin{bmatrix} e^{(1)}_1(\lambda)&
e^{(1)}_2(\lambda)
\end{bmatrix}.
\end{equation}
Similarly, by \eqref{4.7f18}
\begin{equation}\label{4.9f18}
v(\lambda)
\begin{bmatrix} e^{(2)}_2(\lambda)&
e^{(2)}_1(\lambda)
\end{bmatrix}
\begin{bmatrix}F_1\\F_2\end{bmatrix}(v(\lambda))
=f(\lambda).
\end{equation}

It is tempting to make the change of
variable $\lambda\to-\bar\lambda$,
$\lambda\in \bbR$,
in \eqref{4.20d}, \eqref{4.9f18}  and to write
\begin{equation}\label{4.7f15}
v(\lambda)
\begin{bmatrix} (e^{(2)}_2)^-(\lambda)&
(e^{(2)}_1)^-(\lambda)
\end{bmatrix}
\theta(v(\lambda))
=
\begin{bmatrix} (e^{(1)}_1)^-(\lambda)&
(e^{(1)}_2)^-(\lambda)
\end{bmatrix},
\end{equation}
and
\begin{equation}\label{4.11f18}
v(\lambda)
\begin{bmatrix} (e^{(2)}_2)^-(\lambda)&
(e^{(2)}_1)^-(\lambda)
\end{bmatrix}
\begin{bmatrix}F_1\\F_2\end{bmatrix}(v(\lambda))
=f^-(\lambda).
\end{equation}
However,  to suceed with this plan, we need to prove that $\theta (\zeta)$
has an analytic continuation in $\bbC\setminus\bbD$ etc.
That is why we prefer to consider a dual node given by the diagram
\begin{equation}\label{18f15}
\begin{array}{ccc}
K\oplus E_1
&\stackrel{\bar v}{\longrightarrow}& K\oplus E_2\\
\Big\downarrow\rlap{--} & &
\Big\downarrow\rlap{--}\\
K^-\oplus E_1^-
&\stackrel{\bar v}{\longrightarrow}& K^-\oplus E_2^-
\end{array}
\end{equation}
The characteristic matrix-valued function remains the same since we choose basis
in $E_{1,2}^-$ as the image of the basis in $E_{1,2}$. Then 
we obtain \eqref{4.7f15} and \eqref{4.11f18}  simply repeating the arguments 
from \eqref{4.20d} and \eqref{4.9f18}.
Hence
\begin{equation}\label{4.13f18}
v(\lambda)\begin{bmatrix} 
(e^{(2)}_2)^-&
(e^{(2)}_1)^-\\
e^{(2)}_2&
e^{(2)}_1
\end{bmatrix}(\lambda)
\begin{bmatrix}\theta_{11}&\theta_{12}\\
\theta_{21}&\theta_{22}\end{bmatrix}(v(\lambda))
=
\begin{bmatrix} 
(e^{(1)}_1)^-&
(e^{(1)}_2)^-\\
e^{(1)}_1&
e^{(1)}_2
\end{bmatrix}(\lambda),
\end{equation}
and
\begin{equation}\label{4.14f18}
v(\lambda)\begin{bmatrix} 
(e^{(2)}_2)^-&
(e^{(2)}_1)^-\\
e^{(2)}_2&
e^{(2)}_1
\end{bmatrix}(\lambda)
\begin{bmatrix}F_{1}\\
F_{2}\end{bmatrix}(v(\lambda))
=
\begin{bmatrix} 
f^-\\f
\end{bmatrix}(\lambda).
\end{equation}

\begin{lemma}
Both $\det\begin{bmatrix} 
 (e^{(2)}_2)^-& (e^{(2)}_1)^-\\
e^{(2)}_2&
e^{(2)}_1
\end{bmatrix}(\lambda)$ and $\theta_{22}(\lambda)$ do not
vanish identically. Furthermore,
\begin{equation}
\theta_{22}(\lambda)
\det\begin{bmatrix} 
(e^{(2)}_2)^-& (e^{(2)}_1)^-\\
e^{(2)}_2&
e^{(2)}_1 
\end{bmatrix}(\lambda)=-
\frac i{s(\lambda)}\left(\frac 1{v(\lambda)}\right)'.
\end{equation}
In particular, the characteristic matrix-valued function $\theta(\zeta)$
and the map $K_{\scp}(x)\to K_{\theta}$
are well-defined by \eqref{4.13f18}, \eqref{4.14f18}
in terms of the reproducing kernels.
\end{lemma}
\begin{proof}
This follows from an obvious consequence of \eqref{4.13f18}
\begin{equation}\label{4.10f15}
v(\lambda)\begin{bmatrix} 
(e^{(2)}_2)^-&
(e^{(2)}_1)^-\\
e^{(2)}_2&
e^{(2)}_1
\end{bmatrix}(\lambda)
\begin{bmatrix}1&-\theta_{12}\\
0&-\theta_{22}\end{bmatrix}(v(\lambda))
=
\begin{bmatrix} 
v(e^{(2)}_2)^-&
-(e^{(1)}_2)^-\\
v e^{(2)}_2&
-e^{(1)}_2
\end{bmatrix}(\lambda),
\end{equation}
and \eqref{wif2} that says
\begin{equation*}
\det\begin{bmatrix} 
v(e^{(2)}_2)^-&
-(e^{(1)}_2)^-\\
ve^{(2)}_2&
-e^{(1)}_2
\end{bmatrix}(\lambda)= i \frac {v'(\lambda)}{s(\lambda)}.
\end{equation*}
\end{proof}

By \eqref{4.14f24}, we have
for the transfer matrix 
\begin{equation}\label{8.20d}
\mathfrak A_x(\lambda^2)=
\begin{bmatrix}1&-\theta_{12}\\
0&-\theta_{22}\end{bmatrix}^{-1}
\begin{bmatrix}\theta_{11}&0\\
\theta_{21}&-1\end{bmatrix}(v(\lambda)).
\end{equation}
The map from $K_{\theta}$ to the corresponding
de Branges space $\mathcal H(\fA)$ \cite[Sect. 28]{dB} is of the form
\begin{equation}\label{8.20d1}
\begin{bmatrix}A_{1}\\
A_{2}\end{bmatrix}(\lambda^2)=
\begin{bmatrix}1&-\theta_{12}\\
0&-\theta_{22}\end{bmatrix}^{-1}
\begin{bmatrix}F_{1}\\
F_{2}\end{bmatrix}(v(\lambda)).
\end{equation}
Combining \eqref{4.11f15} with
\begin{equation}\label{4.11f15}
v(\lambda)\begin{bmatrix} 
(e^{(2)}_2)^-& (e^{(2)}_1)^-\\
e^{(2)}_2&
e^{(2)}_1
\end{bmatrix}(\lambda)
\begin{bmatrix}\theta_{11}&0\\
\theta_{21}&-1\end{bmatrix}(v(\lambda))
=\begin{bmatrix} 
(e^{(1)}_1)^-&
-v(e^{(2)}_1)^-\\
e^{(1)}_1&
-ve^{(2)}_1
\end{bmatrix}(\lambda),
\end{equation}
we have
\begin{equation}\label{7.20d}
\begin{bmatrix} 
v(e^{(2)}_2)^-&
-(e^{(1)}_2)^-\\
v e^{(2)}_2&
-e^{(1)}_2
\end{bmatrix}(\lambda)
\mathfrak A_x(\lambda^2)
=\begin{bmatrix} 
(e^{(1)}_1)^-&
-v(e^{(2)}_1)^-\\
e^{(1)}_1&
-ve^{(2)}_1
\end{bmatrix}(\lambda),
\end{equation}
and
\begin{equation}\label{4.21f18}
\begin{bmatrix} 
v(e^{(2)}_2)^-&
-(e^{(1)}_2)^-\\
v e^{(2)}_2&
-e^{(1)}_2
\end{bmatrix}(\lambda)
\begin{bmatrix} A_1\\ A_2\
\end{bmatrix}(\lambda^2)
=
\begin{bmatrix} 
f^-\\ f
\end{bmatrix}(\lambda).
\end{equation}

Note that the condition $I-\theta \theta^*\ge 0$ is the same as
\begin{equation*}
\begin{bmatrix}1&-\theta_{12}\\
0&-\theta_{22}\end{bmatrix}
\begin{bmatrix}1&0\\
0&-1\end{bmatrix}
\begin{bmatrix}1&-\theta_{12}\\
0&-\theta_{22}\end{bmatrix}^*-
\begin{bmatrix}\theta_{11}&0\\
\theta_{21}&-1\end{bmatrix}
\begin{bmatrix}1&0\\
0&-1\end{bmatrix}
\begin{bmatrix}\theta_{11}&0\\
\theta_{21}&-1\end{bmatrix}^*\ge 0.
\end{equation*}
That is, the transformation \eqref{8.20d} maps
contractive matrices into $j$-contractions,
$$j-\fA j\fA^*\ge 0,\quad 
j:=\begin{bmatrix}1&0\\0&-1\end{bmatrix},
$$
although this was clear from the definition of $j$-node \eqref{4.8f24}.
 
\subsection{de Branges' Theorem}

\begin{theorem} For every $x>0$, 
\begin{equation}\label{4.29m1}
\mathfrak A_x(\lambda^2)=
\frac{-is(\lambda)}{v'(\lambda)}
\begin{bmatrix} 
-e^{(1)}_2&
(e^{(1)}_2)^-\\
-v e^{(2)}_2&
v(e^{(2)}_2)^-
\end{bmatrix}(\lambda)
\begin{bmatrix} 
(e^{(1)}_1)^-&
-v(e^{(2)}_1)^-\\
e^{(1)}_1&
-ve^{(2)}_1
\end{bmatrix}(\lambda)
\end{equation}
is 
an entire matrix-valued function of $\lambda^2$ and 
\begin{equation}\label{4.30m1}
\mathcal H(\fA_x)=\left\{
\begin{bmatrix} A_1\\ A_2\
\end{bmatrix}(\lambda^2)=
\frac{-is(\lambda)}{v'(\lambda)}
\begin{bmatrix} 
-e^{(1)}_2&
(e^{(1)}_2)^-\\
-v e^{(2)}_2&
v(e^{(2)}_2)^-
\end{bmatrix}(\lambda)
\begin{bmatrix} 
f^-\\ f
\end{bmatrix}(\lambda),\  f\in K_\scp(x)
\right\},
\end{equation}
is the de Branges space of entire functions \cite[Sect. 28]{dB}.
\end{theorem}
The proof is omitted.
 
We point out that for all $x$ the $x$-depending matrix in the RHS
of \eqref{4.29m1}
 meets the following normalization condition
\begin{equation}
(ve_2^{(2)})(\lambda_0)=0,\quad
e_2^{(1)}(\lambda_0)>0.
\end{equation}

As the result
we get a family of $2\times 2$ $j$-contractive 
matrix-valued functions with a certain normalization at $\lambda_0$. 
The family is monotonic in $x$, and every matrix
is an entire function in $\lambda^2$ of the zero mean type
(concerning the corollary of the last condition see \cite{dB}, Theorem 39).  
According to de Branges' Theorem \cite[Sect. 36, 37]{dB}, Theorem 
37, such a family 
can be included in the chain
\begin{equation}\label{9.16d}
j\frac {d}{d t}\mathfrak A(\lambda^2,t)
=
\left\{i\lambda^2\begin{bmatrix} \alpha&\beta\\ \bar\beta&\alpha
\end{bmatrix}(t)
+\begin{bmatrix} 0&-\beta\\ \bar\beta&0
\end{bmatrix}(t)
\right\} \mathfrak A(\lambda^2,t), 
\ |\beta|=\alpha,
\end{equation}
such that $\mathfrak A_x(\lambda^2)=\mathfrak A(\lambda^2,t_x)$,
where $x=x(t)$ is a monotonic function.
Here we choose $\lambda_0^2=i$ as the normalization point.


\subsection{Parameters of the system in terms of reproducing kernels}
\begin{theorem} For the system \eqref{9.16d}
\begin{equation}\label{4.24f21}
\frac{\beta}{\alpha}=-
\frac{s(\lambda_0)}{s(-\bar\lambda_0)}\frac
{d\,\hat k(\lambda_0,-\bar\lambda_0)}{d\,\hat
k(-\bar\lambda_0,-\bar\lambda_0)}.
\end{equation}
\end{theorem}
\begin{proof} Set
\begin{equation}\label{4.25f21}
\mathcal E=\begin{bmatrix}
-e^{(1)}_2&(e^{(1)}_2)^-\\
-ve^{(2)}_2& v(e^{(2)}_2)^-
\end{bmatrix}=\frac{1}{\sqrt{k(\lambda_0,\lambda_0)}}
\begin{bmatrix}
-k_{\lambda_0}&k^-_{\lambda_0}\\
-vk_{-\bar\lambda_0}&vk^-_{-\bar\lambda_0}
\end{bmatrix}.
\end{equation}
We note that due to \eqref{4.29m1}
$
j\dot{\fA}\fA^{-1}=j\dot{\mathcal E}\mathcal E^{-1}.
$
In particular, for $\lambda_0^2=i$
\begin{equation}\label{4.26f21}
j\dot{\mathcal E}(\lambda_0)\mathcal E^{-1}(\lambda_0)
=-\begin{bmatrix} \alpha&2\beta\\ 0&\alpha
\end{bmatrix}.
\end{equation}
On the other hand,
using \eqref{2.1f1}, \eqref{2.1af1}, we have
\begin{equation}\label{4.27f21}
{\mathcal E}(\lambda_0)=\begin{bmatrix}
\tau& a\tau\\ 0&\tau^{-1}
\end{bmatrix}
\begin{bmatrix} -1&0\\0&C
\end{bmatrix}
\end{equation}
with a  constant $C$ and
\begin{equation}\label{4.28f21}
\tau:=\sqrt{k(\lambda_0,\lambda_0)},
\quad a:=s^2(\lambda_0)(2\Im\lambda_0)^2 
\hat k(\lambda_0,-\bar\lambda_0).
\end{equation}
In these notations,
\begin{equation*}
j\dot{\mathcal E}(\lambda_0)\mathcal E^{-1}(\lambda_0)
=\begin{bmatrix} \frac{\dot\tau}{\tau}&\dot a\tau^2\\ 0
&\frac{\dot\tau}{\tau}
\end{bmatrix}.
\end{equation*}
Comparing this with \eqref{4.26f21} we get
$\frac{\beta}{\alpha}=-\frac{da}{d(\tau^{-2})}$. Since
$$
\tau^{-2}=s(\lambda_0)s(-\bar\lambda_0) (2\Im\lambda_0)^2
\hat k(-\bar\lambda_0,-\bar\lambda_0),
$$
by \eqref{4.28f21} we have \eqref{4.24f21}.
\end{proof}

\section{de Branges system and Sturm-Liouville equation}
\label{s5}
In this section, we rewrite the results of the previous sections for a 
particular case of the Sturm-Liouviille equation.
Let
\begin{equation}\label{schtli1}
L y=-y''+q y
\end{equation}
be a self-adjoint operator acting on $L^2(\bbR)$,
and let
\begin{equation}\label{schtli2}
u=\frac{L-\bar\lambda_0^2}{L-\lambda_0^2},
\quad \Re\lambda_0>0, \Im\lambda_0>0,
\end{equation}
be its Cayley transform.
\begin{lemma} Let $e^{\pm}(x,\lambda)\in L^2(\bbR_{\pm})$,
$\Vert e^{\pm}(x,\lambda)\Vert=1$,
be such that
\begin{equation}\label{schtli3}
-\frac{d^2}{dx^2}e^{\pm}(x,\lambda)+q(x) e^{\pm}(x,\lambda)
=\lambda^2 e^{\pm}(x,\lambda),
\quad x\in\bbR_{\pm}, \ \Im\lambda>0, \Re\lambda\not=0.
\end{equation}
Then for all $f_+\in L^2(\bbR_+)$
\begin{equation}\label{schtli4}
u f_+=C e^-(x,\lambda_0)\langle f_+,e^+(x,-\bar\lambda_0)\rangle
+g_+,
\end{equation}
where $C=C(u)$ and $g_+\in L^2(\bbR_+)$. 
\end{lemma}
\begin{proof}
Let
\begin{equation*}
(u f_+)(x)=\begin{cases} g_-(x),& x\in\bbR_-\\
g_+(x), & x\in\bbR_+
\end{cases},
\end{equation*}
or, what is the same,
\begin{equation*}
\left(\frac{\lambda_0^2-\bar\lambda_0^2}{L-\lambda_0^2}
f_+\right)(x)=\begin{cases} g_-(x),& x\in\bbR_-\\
g_+(x)-f_+(x), & x\in\bbR_+
\end{cases}.
\end{equation*}
Therefore,
\begin{equation*}
-g_-''+q g_-=\lambda_0^2 g_-,\quad x\in\bbR_-,
\end{equation*}
and $g_-\in L^2(\bbR_-)$. That is,
\begin{equation*}
g_-(x)= C(f_+)e^{-}(x,\lambda_0).
\end{equation*}
Thus we get
\begin{equation*}
u f_+=C(f_+) e^-(x,\lambda_0)
+g_+.
\end{equation*}

For $C(f_+)$ we have
\begin{equation*}
C(f_+)=\langle uf_+, e^-(x,\lambda_0)\rangle=
\langle f_+, u^*e^-(x,\lambda_0)\rangle.
\end{equation*}
Now we are looking at
\begin{equation*}
u^*e^-(x,\lambda_0)=
\left(I+\frac{\bar\lambda_0^2-\lambda_0^2}
{L-\bar\lambda_0^2}\right)e^-(x,\lambda_0)=
\begin{cases}
h_-(x),\ x\in\br_-\\
h_+(x),\ x\in\br_+
\end{cases},
\end{equation*}
or
\begin{equation*}
(\bar\lambda_0^2-\lambda_0^2)e^-(x,\lambda_0)=
\begin{cases}
-\tilde h''_-+q\tilde h_- -\bar\lambda_0^2\tilde h_-,\ x\in\br_-\\
-\tilde h''_+ +q\tilde h_+ -\bar\lambda_0^2\tilde h_+,\ x\in\br_+
\end{cases},
\end{equation*}
where $\tilde h=h-e^-(x,\lambda_0)$.
This implies
\begin{equation*}
\tilde h=
\begin{cases}
-e^-(x,\lambda_0)+ C_1e^-(x,-\bar\lambda_0),\ x\in\br_-\\ 
C_2 e^+(x,-\bar\lambda_0),\ x\in\br_+
\end{cases},
\end{equation*}
where $C_1, C_2$ are defined by the conditions
$$
\tilde h(-0)=\tilde h(+0),\quad
\tilde h'(-0)=\tilde h'(+0).
$$
Notice that the equality  $C_2=0$ contradicts the linear independence
of $e^-(x,\lambda_0)$ and $e^-(x,-\bar\lambda_0)$.

Hence,
\begin{equation}\label{schtli5}
 u^* e^-(x,\lambda_0)=
\tilde h + e^-(x,\lambda_0)=
\begin{cases} 
C_1e^-(x,-\bar\lambda_0),\ x\in\br_-\\ 
C_2 e^+(x,-\bar\lambda_0),\ x\in\br_+
\end{cases},
\end{equation}
and \eqref{schtli4} is proved with
$$
C(f_+)=\langle f_+, C_2 e^+(x,-\bar\lambda_0)\rangle.
$$

\end{proof}
\begin{corollary} The operator
u acts from $L^2(\bbR_+)\oplus\{ e^-(x,-\bar\lambda_0)\}$
to $L^2(\bbR_+)\oplus\{ e^-(x,\lambda_0)\}$.
\end{corollary}
\begin{proof} Similarly to \eqref{schtli4},
\begin{equation}\label{schtli6}
u f_-=C e^+(x,\lambda_0)\langle f_-,e^-(x,-\bar\lambda_0)\rangle
+g_-.
\end{equation}
That is, $ uf_-=g_ -\in L^2(\bbR_-)$ if
$f_-\perp e^-(x,-\bar\lambda_0)$.
Moreover,
$(uf_-)$ is orthogonal to $e^-(x,\lambda_0)$  by \eqref{schtli5} in this case.
\end{proof}

\begin{corollary} For every $x_0>0$, the operator
u acts from 
\begin{equation}\label{schtli7}
(L^2[0,x_0]\oplus\{ e^-(x,-\bar\lambda_0)\})
\oplus\{ e_{x_0}^+(x,-\bar\lambda_0)\}
\end{equation}
to 
\begin{equation}\label{schtli8}
(L^2[0,x_0]\oplus\{ e_{x_0}^+(x,\lambda_0)\})\oplus\{
e^-(x,\lambda_0)\}
\end{equation}
where $e_{x_0}^+(x,\lambda)\in L^2[x_0,\infty)$ is the
normalized solution of \eqref{schtli3}.
\end{corollary}

\begin{theorem}
The transfer matrix of the unitary node \eqref{schtli7},
\eqref{schtli8} is of the form
\begin{equation}\label{schtli9t}
\fA_{x_0}(\lambda^2)=
\begin{bmatrix} e_{x_0}^+(x_0,\lambda_0)& 
-e_{x_0}^+(x_0,-\bar\lambda_0)\\
\dot e_{x_0}^+(x_0,\lambda_0)& 
-\dot e_{x_0}^+(x_0,-\bar\lambda_0)
\end{bmatrix}^{-1}
\fB_{x_0}(\lambda^2)
\begin{bmatrix} -e^-(0,-\bar\lambda_0)& e^-(0,\lambda_0)\\
-\dot e^-(0,-\bar\lambda_0)&\dot e^-(0,\lambda_0)
\end{bmatrix},
\end{equation}
where
\begin{equation}
\fB_x(\lambda^2)=
\begin{bmatrix} c(x,\lambda)&s(x,\lambda)\\
 \dot c(x,\lambda)&\dot s(x,\lambda)
\end{bmatrix}
\end{equation}
is the standard transfer matrix for equation \eqref{schtli1}
\begin{equation}
\frac{d}{dx}\fB_x(\lambda^2)=
\begin{bmatrix} 0&1\\
 q(x)-\lambda^2& 0
\end{bmatrix}\fB_x(\lambda^2),
\quad \fB_0(\lambda^2)=I.
\end{equation}
\end{theorem}

\begin{proof}
In the block form we have
\begin{equation}\label{schtli9}
u\begin{bmatrix}
\zeta k_{\zeta}\\
\begin{bmatrix} c_1\\d_2
\end{bmatrix}
\end{bmatrix}=
\begin{bmatrix}
 k_{\zeta}\\
\begin{bmatrix} d_1\\c_2
\end{bmatrix}
\end{bmatrix},
\end{equation}
with
\begin{equation}\label{schtli10}
\begin{bmatrix} d_1\\d_2
\end{bmatrix}=
\fA(\zeta)
\begin{bmatrix} c_1\\c_2
\end{bmatrix}.
\end{equation}
In other words,
\begin{equation*}
\begin{split}
\left( I+\frac{\lambda_0^2-\bar\lambda_0^2}{L-\lambda_0^2}
\right)&\{\zeta k_{\zeta} +c_1 e^-(x,-\bar\lambda_0)+
d_2 e_{x_0}^+(x,-\bar\lambda_0)\}\\
=& k_{\zeta} +d_1 e_{x_0}^+(x,\lambda_0)+
c_2 e^-(x,\lambda_0),
\end{split}
\end{equation*}
or
\begin{equation}\label{schtli11}
\begin{split}
\frac{\lambda_0^2-\bar\lambda_0^2}{L-\lambda_0^2}
&\{\zeta k_{\zeta} +c_1 e^-(x,-\bar\lambda_0)+
d_2 e_{x_0}^+(x,-\bar\lambda_0)\}\\
=& (1-\zeta)k_{\zeta} +d_1 e_{x_0}^+(x,\lambda_0)+
c_2 e^-(x,\lambda_0)
-c_1 e^-(x,-\bar\lambda_0)-
d_2 e_{x_0}^+(x,-\bar\lambda_0).
\end{split}
\end{equation}
This means that the RHS of \eqref{schtli11} has the second
derivative and we have on the interval $[0,x_0]$
\begin{equation}\label{schtli12}
-k''_{\zeta}+qk_{\zeta}=\lambda^2 k_{\zeta}.
\end{equation}
for the spectral parameter 
$$
\lambda^2=
\lambda_0^2+\frac{\zeta}{1-\zeta}
(\lambda_0^2-\bar\lambda_0^2).
$$
Above, $\zeta=\frac{\lambda^2-\lambda_0^2}
{\lambda^2-\bar\lambda_0^2}$.

Let
\begin{equation}\label{schtli13}
(1-\zeta) k_{\zeta}=A c(x,\lambda)+B s(x,\lambda).
\end{equation}
 Then the continuity at $x=0$ implies
\begin{equation*}
\begin{bmatrix} A\\B
\end{bmatrix}=
\begin{bmatrix} -e^-(0,-\bar\lambda_0)& e^-(0,\lambda_0)\\
-\dot e^-(0,-\bar\lambda_0)&\dot e^-(0,\lambda_0)
\end{bmatrix}
\begin{bmatrix} c_1\\c_2
\end{bmatrix},
\end{equation*}
and by the continuity at $x=x_0$,
\begin{equation*}
\begin{bmatrix} e_{x_0}^+(x_0,\lambda_0)& 
-e_{x_0}^+(x_0,-\bar\lambda_0)\\
\dot e_{x_0}^+(x_0,\lambda_0)& 
-\dot e_{x_0}^+(x_0,-\bar\lambda_0)
\end{bmatrix}
\begin{bmatrix} d_1\\d_2
\end{bmatrix}=
\begin{bmatrix} c(x_0,\lambda)&s(x_0,\lambda)\\
 \dot c(x_0,\lambda)&\dot s(x_0,\lambda)
\end{bmatrix}
\begin{bmatrix} A\\B
\end{bmatrix}.
\end{equation*}
The theorem is proved.
\end{proof}

We now compute the parameters of the related canonical
system under the chosen normalization.

We start observing that, up to the initial matrix
$\fA_0$, the transfer matrix has the same normalization as the transfer matrix 
\eqref{8.20d} (or \eqref{4.29m1})
in Section 4.
\begin{corollary}
The transfer matrix of unitary node \eqref{schtli7},
\eqref{schtli8} is of the form
\begin{equation*}
\fA_x(\lambda^2)=\tilde{\fA}_x(\lambda^2)\fA_0,
\end{equation*}
where
\begin{equation}\label{17m9}
\tilde{\fA}_{x_0}(\lambda^2)=
\begin{bmatrix} e_{x_0}^+(x_0,\lambda_0)& 
-e_{x_0}^+(x_0,-\bar\lambda_0)\\
\dot e_{x_0}^+(x_0,\lambda_0)& 
-\dot e_{x_0}^+(x_0,-\bar\lambda_0)
\end{bmatrix}^{-1}
\fB_{x_0}(\lambda^2)
\begin{bmatrix} e_{0}^+(0,\lambda_0)& 
-e_{0}^+(0,-\bar\lambda_0)\\
\dot e_{0}^+(0,\lambda_0)& 
-\dot e_{0}^+(0,-\bar\lambda_0)
\end{bmatrix}.
\end{equation}
Therefore, $\tilde{\fA}_{x}(\lambda^2)$ meets the 
normalization:
\begin{equation*}
 (\tilde{\fA}_{x}(\lambda_0^2))_{11}>0,
 \quad (\tilde{\fA}_{x}(\lambda_0^2))_{21}=0
\end{equation*}
for all $x>0$.
\end{corollary}

We use the same notation
as in \eqref{4.27f21}.

\begin{theorem} Let $m_+(\lambda)$ be the Weyl function of operator 
\eqref{schtli1}
and let  
\begin{equation}\label{20m9}
\tilde{\fA}_x(\lambda_0^2)=
\begin{bmatrix}
\tau&a\tau\\ 0&\tau^{-1}
\end{bmatrix}.
\end{equation}
Then
\begin{equation}\label{m9}
\frac{d a}{d(\tau^{-2})}
=\frac{\overline{c(x,\lambda_0)}\,\overline{m_+(\lambda_0)}+
\overline{s(x,\lambda_0)}}
{{c(x,\lambda_0)}\,{m_+(\lambda_0)}+{s(x,\lambda_0)}}.
\end{equation}
\end{theorem}

\begin{proof}
First of all,
\begin{equation}\label{21m9}
\begin{bmatrix}
e^+_{x_0}(x,\lambda)\\
\dot e^+_{x_0}(x,\lambda)
\end{bmatrix}= \fB_{x}
\begin{bmatrix}
m_+(\lambda)\\
1
\end{bmatrix}\rho(x_0),\quad x\ge x_0,
\end{equation}
where $\rho(x_0)$ should be found from the condition
\begin{equation*}
\int_{x_0}^{\infty}|e^+_{x_0}(x,\lambda)|^2\,dx=1.
\end{equation*}
Using
\begin{equation}\label{22m9}
\frac{d}{dx}\{\fB^*_x(\lambda^2)J\fB_x(\lambda^2)\}= -
(\lambda^2-\bar\lambda^2)\fB^*_x(\lambda^2)
\begin{bmatrix} 1\\0\end{bmatrix}
\begin{bmatrix} 1 &0\end{bmatrix}
\fB_x(\lambda^2),
\end{equation}
with $J=\begin{bmatrix}0&1\\-1&0\end{bmatrix},$
we obtain
\begin{equation}\label{23m9}
\begin{split}
\rho^2(x_0)&\int_{x_0}^{\infty}
\begin{bmatrix}
\overline{m_+(\lambda)}&
1
\end{bmatrix}
\fB^*_x(\lambda^2)
\begin{bmatrix} 1\\0\end{bmatrix}
\begin{bmatrix} 1 &0\end{bmatrix}
\fB_x(\lambda^2)
\begin{bmatrix}
m_+(\lambda)\\
1
\end{bmatrix}{dx}\\
=\rho^2(x_0)&\frac
{\begin{bmatrix}
\overline{m_+(\lambda)}&
1
\end{bmatrix}
\fB^*_{x_0}(\lambda^2)J\fB_{x_0}(\lambda^2)
\begin{bmatrix}
{m_+(\lambda)}\\
1
\end{bmatrix}
}
{\lambda^2-\bar\lambda^2}.
\end{split}
\end{equation}
That is,
\begin{equation}\label{24m9}
\rho^2(x_0)=\frac{\lambda^2-\bar\lambda^2}
{\begin{bmatrix}
\overline{m_+(\lambda)}&
1
\end{bmatrix}
\fB^*_{x_0}(\lambda^2)J\fB_{x_0}(\lambda^2)
\begin{bmatrix}
{m_+(\lambda)}\\
1
\end{bmatrix}}.
\end{equation}
In particular
 \begin{equation}\label{25m9}
\rho^2(0)=-\frac{\lambda^2-\bar\lambda^2}
{m_+(\lambda)-
\overline{m_+(\lambda)}}.
\end{equation}
Since for $x\ge x_0$
\begin{equation}\label{26m9}
\fB_x(\lambda_0^2)
\begin{bmatrix}
e^+_{0}(0,\lambda_0)\\
\dot e^+_{0}(0,\lambda_0)
\end{bmatrix}= \fB_{x}(\lambda_0^2)
\begin{bmatrix}
m_+(\lambda_0)\\
1
\end{bmatrix}\rho(0)
=
\begin{bmatrix}
e^+_{x_0}(x,\lambda_0)\\
\dot e^+_{x_0}(x,\lambda_0)
\end{bmatrix}\frac{\rho(0)}{\rho(x_0)},
\end{equation}
we get for the first column of the matrix 
 $\tilde{\fA}_x(\lambda_0^2)$ \eqref{17m9}
 \begin{equation*}
\begin{bmatrix} e_{x_0}^+(x_0,\lambda_0)& 
-e_{x_0}^+(x_0,-\bar\lambda_0)\\
\dot e_{x_0}^+(x_0,\lambda_0)& 
-\dot e_{x_0}^+(x_0,-\bar\lambda_0)
\end{bmatrix}^{-1}
\begin{bmatrix}
e^+_{x_0}(x_0,\lambda_0)\\
\dot e^+_{x_0}(x_0,\lambda_0)
\end{bmatrix}\frac{\rho(0)}{\rho(x_0)}=
\begin{bmatrix}1\\0\end{bmatrix}
\frac{\rho(0)}{\rho(x_0)}.
\end{equation*}
Therefore, we deduce from \eqref{20m9} that $\tau=\frac{\rho(0)}{\rho(x_0)}$ 
and, recalling
\eqref{24m9}, \eqref{25m9}, we come to
\begin{equation}\label{24m9tau}
\tau^{-2}=-\frac{m_+(\lambda_0)-\overline{m_+(\lambda_0)}}
{\begin{bmatrix}
\overline{m_+(\lambda_0)}&
1
\end{bmatrix}
\fB^*_{x_0}(\lambda_0^2)J\fB_{x_0}(\lambda_0^2)
\begin{bmatrix}
{m_+(\lambda_0)}\\
1
\end{bmatrix}}.
\end{equation}

To compute $a\tau$, we proceed as
\begin{equation}\label{27m9}
\begin{split} a\tau \Delta=& 
-\begin{bmatrix}
-\dot e^+_{x_0}(x_0,-\bar\lambda_0)&
e^+_{x_0}(x_0,-\bar\lambda_0)
\end{bmatrix}
\fB_{x_0}(\lambda_0^2)
\begin{bmatrix}
\overline{m_+(\lambda_0)}\\
1
\end{bmatrix}\rho(0)
\\
=& 
-\begin{bmatrix}
e^+_{x_0}(x_0,-\bar\lambda_0)&
\dot e^+_{x_0}(x_0,-\bar\lambda_0)
\end{bmatrix}J
\fB_{x_0}(\lambda_0^2)
\begin{bmatrix}
\overline{m_+(\lambda_0)}\\
1
\end{bmatrix}\rho(0)
\\
=&-
\rho(x_0)\begin{bmatrix}
\overline{m_+(\lambda_0)}&
1
\end{bmatrix}\fB^*_{x_0}(\lambda_0^2)J
\fB_{x_0}(\lambda_0^2)
\begin{bmatrix}
\overline{m_+(\lambda_0)}\\
1
\end{bmatrix}\rho(0),
\end{split}
\end{equation}
where 
\begin{equation}\label{28m9}
\begin{split}
\Delta=&\det\begin{bmatrix} e_{x_0}^+(x_0,\lambda_0)& 
-e_{x_0}^+(x_0,-\bar\lambda_0)\\
\dot e_{x_0}^+(x_0,\lambda_0)& 
-\dot e_{x_0}^+(x_0,-\bar\lambda_0)
\end{bmatrix}\\
=&\begin{bmatrix}
e^+_{x_0}(x_0,-\bar\lambda_0)&
\dot e^+_{x_0}(x_0,-\bar\lambda_0)
\end{bmatrix}J\begin{bmatrix}
e^+_{x_0}(x_0,\lambda_0)\\
\dot e^+_{x_0}(x_0,\lambda_0)
\end{bmatrix}\\
=&\rho(x_0)\begin{bmatrix}
\overline{m_+(\lambda_0)}&
1
\end{bmatrix}\fB^*_{x_0}(\lambda_0^2)J
\fB_{x_0}(\lambda_0^2)
\begin{bmatrix}
{m_+(\lambda_0)}\\
1
\end{bmatrix}\rho(x_0).
\end{split}
\end{equation}
Combining \eqref{27m9} and \eqref{28m9}, we obtain
\begin{equation}\label{29m9}
a=-\frac{
\begin{bmatrix}
\overline{m_+(\lambda_0)}&
1
\end{bmatrix}
\fB^*_{x_0}(\lambda_0^2)J\fB_{x_0}(\lambda_0^2)
\begin{bmatrix}
\overline{m_+(\lambda_0)}\\1
\end{bmatrix}}{
\begin{bmatrix}
\overline{m_+(\lambda_0)}&
1
\end{bmatrix}\fB^*_{x_0}(\lambda_0^2)J
\fB_{x_0}(\lambda_0^2)
\begin{bmatrix}
{m_+(\lambda_0)}\\ 1\end{bmatrix}}.
\end{equation}
Using \eqref{22m9} and the Wronskian identity for $\fB_x(\lambda)$
we get \eqref{m9} from \eqref{24m9tau} and \eqref{29m9}
by a direct computation.
\end{proof}

\section{Appendix 1. An example}
\label{s6}
In this section we give an example which shows that the class of canonical 
systems discussed in Section \ref{s4} is larger than the class of 
Sturm-Liouville equations from Section \ref{s5}. We will see
that generally 
\begin{equation}\label{ee6}
H^2_{\scp}\not=\hat H^2_{\scp},
\end{equation}
although always $H^2_{\scp}\subset\hat H^2_{\scp}$ and we will also discuss some 
other
interesting phenomena.

{\it Throughout this section
we set $s_+=\frac{\lambda}{\lambda+i}$ and $\nu_+=0$.}

First, we prove \eqref{ee6}. Since
$$
\langle (s_+f)(\lambda),-f(-\bar\lambda)\rangle=0
$$
for all $f\in H^2$, we get
$||f||_{\scp}=||f||$. Therefore, in this case $H^2_{\scp}$ coincides with
the standard $H^2$.

On the other hand, we have $s(\lambda)=\frac{i}{\lambda+i}$,
so $s\cdot 1\in H^2$. Let us check that
$1\in L^2_{\scp}$. This follows from the identity
\begin{equation}\label{0.5j}
\begin{bmatrix}1& -1
\end{bmatrix}
\begin{bmatrix}1&  \overline{
\frac{\lambda}{\lambda+i}}\\
\frac{\lambda}{\lambda+i}&1
\end{bmatrix}
\begin{bmatrix}1\\ -1
\end{bmatrix}=\frac{2}{|\lambda+i|^2}.
\end{equation}
Hence, by the definition of $\hat H^2_{\scp}$
a constant function belongs to this space, but of course
$1\not\in H^2$ and \eqref{ee6} is proved. 

The above conclusion can be sharpened. Using
\begin{equation}\label{1.10j}
\frac 1 2\left\langle
\begin{bmatrix}1&  
\overline{s_+(\lambda)}\\
s_+(\lambda)&1
\end{bmatrix}
\begin{bmatrix}1\\ -1
\end{bmatrix},
\begin{bmatrix}f(\lambda)\\ -f(-\bar\lambda)
\end{bmatrix}
\right\rangle
=\langle s_+(\lambda)-1,-f(-\bar\lambda)\rangle=0,
\end{equation}
for all $f\in H^2$, we get
 that
$1$ is orthogonal to $H^2_{\scp}
\subset\hat H^2_{\scp}$. Actually we have the following orthogonal
decomposition
\begin{equation}\label{1.5j}
\hat H^2_{\scp}=\{1\}\oplus H^2_{\scp}
=\{1\}\oplus H^2.
\end{equation}

This implies that  the reproducing kernel
of
$\hat H^2_{\scp}$ is 
\begin{equation}\label{2.5j}
\hat k(\lambda,\lambda_0)=\hat k_{\scp}(\lambda,\lambda_0)=
\frac 1{||1||^2_{\scp}}+\frac i{\lambda-\bar\lambda_0},
\end{equation}
and, by \eqref{0.5j}, $||1||^2_{\scp}=\frac 1 2$.

Now we show that the property
$H^2_{\scp}(x)\not=\hat H^2_{\scp}(x)$ is not $x$-invariant.
Namely, $H^2_{\scp}(x)=\hat H^2_{\scp}(x)$ for $x>0$ despite \eqref{ee6} for 
$x=0$.

\begin{lemma} Let $s_+=\frac{\lambda}{\lambda+i}$, $\nu_+=0$.
Then
$H^2_{\scp}(x)=\hat H^2_{\scp}(x)$ for all $x>0$.
\end{lemma}

\begin{proof}
Notice that
$$
\hat H^2_{\scp}(x)=e^{i\lambda x}
\hat H^2_{\{s_+e^{2i\lambda x},\nu_+e^{2i\lambda x}\}},
$$
and $H^2_{\scp}(x)=e^{i\lambda x} H^2$. So we have to show
that $\hat H^2_{\{s_+e^{2i\lambda x},\nu_+e^{2i\lambda x}\}}
= H^2$.

By definition $f\in\hat H^2_{\{s_+e^{2i\lambda x},\nu_+e^{2i\lambda
x}\}}$ means that
\begin{equation}\label{2.10j}
\begin{split}
&\frac{i}{\lambda+i}f(\lambda)\in H^2,\\
&e^{2i\lambda x}\frac{\lambda}{\lambda+i}f(\lambda)
-f(-\bar\lambda)\in L^2.
\end{split}
\end{equation}
We have to prove that $f\in H^2$. 

Let
$g(\lambda)=\frac{i}{\lambda+i}f(\lambda)$. Conditions
\eqref{2.10j} can be easily transformed into
\begin{equation*}
\lambda\{e^{2i\lambda x}g(\lambda)
+g(-\bar\lambda)\}\in L^2
\end{equation*}
with $g\in H^2$. Let $G$ denote the Fourier transform of $g$. Obviously, $G\in 
L^2(\br_+)$ since $g\in H^2$. In these terms we have
\begin{equation}\label{3.10j}
\{G(2x+t)
+G(-t)\}'\in L^2.
\end{equation}
Since the supports of the functions
$G(2x+t)$ and
$G(-t)$ do not intersect,  we get from
\eqref{3.10j} that $G'(t)\in L^2$.
Therefore $\lambda g(\lambda)\in L^2$, and, consequently, 
$f$ belongs to $L^2$ and, in fact, to $H^2$.
\end{proof}

\begin{corollary}\label{c4.2}
Let $s_+=\frac{\lambda}{\lambda+i}$, $\nu_+=0$.
Then
$H^2_{\scp}(x)=\hat H^2_{\scp}(x)$ for all $x<0$.
\end{corollary}
\begin{proof} We only have to mention that $s_-=s_+$ in our case
and to use Theorem \ref{t1.3}.
\end{proof}
\begin{corollary} Let $s_+=\frac{\lambda}{\lambda+i}$, $\nu_+=0$.
Then
\begin{equation}\label{4.10j}
\lim_{x\to-0}H^2_{\scp}(x)=\hat H^2_{\scp}(0).
\end{equation}
\end{corollary}
\begin{proof} Obviously,
$\lim_{x\to x_0+0}H^2_{\scp}(x)= H^2_{\scp}(x_0)$
for $x_0\ge0$.
Therefore by $s_-=s_+$ and the duality
stated in Theorem \ref{t1.3},
$$
\lim_{x\to x_0-0}\hat H^2_{\scp}(x)= \hat H^2_{\scp}(x_0)
$$
for $x_0\le 0$. Finally, we use Corollary \ref{c4.2}
$$
\lim_{x\to -0} H^2_{\scp}(x)=
\lim_{x\to -0}\hat H^2_{\scp}(x)= \hat H^2_{\scp}(0).
$$
\end{proof}

This means, in particular, that
the canonical system related to the given scattering data is not a 
Sturm-Liouville equation.

Indeed, let $\mathfrak A(x_0,x_1;\lambda^2)$, $x_0<x_1$, be the
transfer matrix \eqref{8.20d} (or \eqref{4.29m1}). Recalling \eqref{7.20d}, we 
write
\begin{equation*}
\begin{bmatrix}
v e^{(2)}_2(x_1)& e^{(1)}_2(x_1)
\end{bmatrix}=
\begin{bmatrix}
e^{(1)}_1(x_0)& v e^{(2)}_1(x_0)
\end{bmatrix}
\mathfrak A(x_0,x_1;\lambda^2).
\end{equation*}
We also introduce  $\mathfrak B(x_0,x_1;\lambda^2)$ by
\begin{equation}\label{5.10j}
\begin{bmatrix}
v e^{(2)}_2(x_1)& e^{(1)}_2(x_1) 
\end{bmatrix}=
\begin{bmatrix}
v e^{(2)}_2(x_0)& e^{(1)}_2(x_0)
\end{bmatrix}
\mathfrak B(x_0,x_1;\lambda^2),
\end{equation}
so that we have the chain rule
$$
\mathfrak A(x_0,x_2;\lambda^2)=
\mathfrak A(x_0,x_1;\lambda^2)\mathfrak B(x_1,x_2;\lambda^2).
$$
Fix $x_0<0$, put $x_2=0$,  and let
$$
\mathfrak B(\lambda^2)=\lim_{x_1\to -0}\mathfrak B(x_1,0;\lambda^2).
$$
Using
\eqref{5.10j} and \eqref{4.10j} we have
\begin{equation}\label{1.11j}
\begin{bmatrix}
v e^{(2)}_2(x_1)& e^{(1)}_2(x_1) 
\end{bmatrix}=
\begin{bmatrix}
v \hat e^{(2)}_2(x_1)& \hat e^{(1)}_2(x_1) 
\end{bmatrix}
\mathfrak B(\lambda^2),
\end{equation}
where (see \eqref{1.20d2}, \eqref{2.20d})
$$
\hat e^{(1)}_2(\l)=\frac{\hat k_{\scp}(\lambda,\lambda_0;x)}{||\hat 
k_{\scp}(\lambda,\lambda_0;x)||},\quad 
\hat e^{(2)}_2(\l)=\frac{\hat k_{\scp}(\lambda,-\bar\lambda_0;x)}{||\hat 
k_{\scp}(\lambda,-\bar\lambda_0;x)||}.
$$
Thus $\mathfrak B(\lambda^2)$ is a {\it non-trivial divisor} of
$\mathfrak A(x_0,x;\lambda^2)$ for $x\ge 0$. The chain 
$\mathfrak A(x_0,x;\lambda^2)$ is not even
continuous in $x$,
$$
\lim_{x\to -0}\mathfrak A(x_0,x;\lambda^2)
\not =\mathfrak A(x_0,0;\lambda^2).
$$

Of course, we can get an explicit formula for
$\mathfrak B(\lambda^2)$.
Since $\mathfrak B$ depends on $\lambda^2$,  \eqref{1.11j} implies
\begin{equation*}
 \begin{bmatrix} 
 e^{(2)}_2(\lambda)&e^{(1)}_2(\lambda,\lambda_0)\\
 e^{(2)}(-\lambda)&e^{(1)}_2(-\lambda)
\end{bmatrix}
\begin{bmatrix}v(\lambda)&0\\0&1
\end{bmatrix}
=
\begin{bmatrix}
\hat e^{(2)}_2(\lambda)&\hat e^{(1)}_2(\lambda)\\
\hat e^{(2)}_2(-\lambda)&\hat e^{(1)}_2(-\lambda)
\end{bmatrix}
\begin{bmatrix}v(\lambda)&0\\0&1
\end{bmatrix}
\mathfrak B(\lambda^2),
\end{equation*}
or, with the help of \eqref{2.5j},
\begin{equation}\label{2.11j}
\begin{split}
\sqrt{
\frac{\hat k(\lambda_0,\lambda_0)}{k(\lambda_0,\lambda_0)}}
&\begin{bmatrix} k(\lambda,-\bar\lambda_0)&
 k(\lambda,\lambda_0)\\
 k(-\lambda,-\bar\lambda_0)&
 k(-\lambda,\lambda_0)
\end{bmatrix}
\begin{bmatrix}v(\lambda)&0\\0&1
\end{bmatrix}
\\=&
\begin{bmatrix}\hat k(\lambda,-\bar\lambda_0)&
\hat k(\lambda,\lambda_0)\\
\hat k(-\lambda,-\bar\lambda_0)&
\hat k(-\lambda,\lambda_0)
\end{bmatrix}
\begin{bmatrix}v(\lambda)&0\\0&1
\end{bmatrix}
\mathfrak B(\lambda^2)\\
=&
\left\{ 2 \begin{bmatrix}1&1\\1&1
\end{bmatrix}+
\begin{bmatrix} k(\lambda,-\bar\lambda_0)&
 k(\lambda,\lambda_0)\\
 k(-\lambda,-\bar\lambda_0)&
 k(-\lambda,\lambda_0)
\end{bmatrix}\right\}
\begin{bmatrix}v(\lambda)&0\\0&1
\end{bmatrix}
\mathfrak B(\lambda^2).
\end{split}
\end{equation}
Thus, directly,
\begin{equation*}
\sqrt{1+4\Im\lambda_0}I=\left\{\frac{i}{\Re\lambda_0}
\begin{bmatrix}\lambda^2-\lambda_0^2&\lambda^2-\bar\lambda_0^2\\
-\lambda^2+\lambda_0^2&-\lambda^2+\bar\lambda_0^2
\end{bmatrix}+I
\right\}\mathfrak B(\lambda^2).
\end{equation*}
Note that the determinant of the matrix in curly brackets is
$1+4\Im\lambda_0$, so $\mathfrak B(\lambda^2)$ is indeed an entire
function of $\lambda^2$ (a linear polynomial),
\begin{equation*}
\mathfrak B(\lambda^2)=\frac{1}
{\sqrt{1+4\Im\lambda_0}}\left\{I+\frac{i}{\Re\lambda_0}
\begin{bmatrix}-\lambda^2+\bar\lambda_0^2&-\lambda^2+\bar\lambda_0^2\\
\lambda^2-\lambda_0^2&\lambda^2-\lambda_0^2
\end{bmatrix}
\right\}.
\end{equation*}

\section{Appendix 2. On a certain sufficient condition}
\label{s7}

\subsection{An extension of $A_2$ in the presence of the mass points }
\begin{theorem} 
Let $E=[-2,2]$ and $X=\{x_k\}$ be a set of points on $\bbR\setminus E$ that 
satisfies the Blaschke conditionin the domain $\bbC\setminus E$. 
Let $\Sigma$ be $n\times n$ matrix-measure supported on $E\cup X$ which 
is absolutely continuos on $E$,
\begin{equation}
d\Sigma(x)=W(x)\,dx,
\end{equation}
moreover, $W(x)^{-1}$ exists for almost all $x\in E$, and 
$\Sigma(x_k)=\Sigma_k$. Define
\begin{equation}(\frak H f)(x)=\lim_{\epsilon\to+0}\int_{E\cup 
X}\frac{d\Sigma(t)f(t)}{t-(x+i\epsilon)},\quad x\in E,
\end{equation}
for smooth vector--functions $f(t)$. Then there exists $Q>0$ such that
\begin{equation}
\int_E(\frak H f)^*(x) W(x)(\frak H f)(x)\,dx\le Q\int_{E\cup X} f^*(t) 
d\Sigma(t) f(t)
\end{equation}
for all such $f$'s if and only if $W$ belongs to matrix $A_2$, and  we have the 
following Carleson type inequality for any vector function 
$f\in L^2(W^{-1})$:
\begin{equation}
\label{carl1}
\sum_{x-k\in X} \langle \Sigma_k f(x_k), f(x_k)\rangle \leq Q \int_{E}\langle 
W^{-1} f,f\rangle dx\,.
\end{equation}
Here $f(x_k) := \int_{E} \frac{f(t)dt}{x_k-t}$.
\end{theorem}

\subsection{On a certain sufficient condition} The following lemmas
are related to attempts to rewrite the $A_2$ condition for the 
spectral density directly in terms of the scattering function.
\begin{lemma} Let 
$$
W=\begin{bmatrix} 1&\bar s_+\\
s_+&1
\end{bmatrix}.
$$
The following conditions are equivalent
\begin{equation}\label{1sos}
\left\langle
W^{-1}P_+W
\begin{bmatrix} f(t)\\
\bar t f(\bar t)
\end{bmatrix},
P_+W
\begin{bmatrix} f(t)\\
\bar t f(\bar t)
\end{bmatrix}\right\rangle\le
Q
\left\langle
W
\begin{bmatrix} f(t)\\
\bar t f(\bar t)
\end{bmatrix},
\begin{bmatrix} f(t)\\
\bar t f(\bar t)
\end{bmatrix}\right\rangle
\end{equation}
for all $f\in L^2_{s_+}\ominus H^2_{s_+}$ and
\begin{equation}\label{2sos}
||f^-||^2\le Q||f^-||^2_{s_-}
\end{equation}
for $f^-(t)\in \hat H^2_{s_-}$.
\end{lemma}
\begin{lemma} If \eqref{2sos} holds, then
$\hat H^2_{s_-}= H^2_{s_-}$, moreover the norm in
$\hat H^2_{s_-}$ is equivalent to the standard $H^2$-norm.
\end{lemma}
\begin{lemma} $\hat H^2_{s_-}= H^2_{s_-}$ implies
$\hat H^2_{s_+}= H^2_{s_+}$
\end{lemma}
\begin{proof}
\begin{equation*}
\begin{split}
\hat H^2_{s_+}=&(L^2_{s_-}\ominus H^2_{s_-})^+ \\
=&(L^2_{s_-}\ominus \hat H^2_{s_-})^+\\
=&L^2_{s_+}\ominus (\hat H^2_{s_-})^+=
H^2_{s_+}.
\end{split}
\end{equation*}
\end{proof}
Nevertheless, we cannot guarantee that the norm
in $H^2_{s_+}$ is equivalent to the $H^2$-norm. Thus in
addition to \eqref{2sos} we have to impose the condition
\begin{equation}\label{3sos}
\langle f,f\rangle\le
Q
\left\langle
W
\begin{bmatrix} f(t)\\
\bar t f(\bar t)
\end{bmatrix},
\begin{bmatrix} f(t)\\
\bar t f(\bar t)
\end{bmatrix}\right\rangle
\end{equation}
for all $f\in H^2$ (this is exactly the condition on equivalence  of the norms).
Obviously, the last inequality is the same as
\begin{equation}\label{4sos}
\left\langle
W P_+
\begin{bmatrix} f(t)\\
\bar t f(\bar t)
\end{bmatrix}, P_+
\begin{bmatrix} f(t)\\
\bar t f(\bar t)
\end{bmatrix}\right\rangle\le
Q
\left\langle
W
\begin{bmatrix} f(t)\\
\bar t f(\bar t)
\end{bmatrix},
\begin{bmatrix} f(t)\\
\bar t f(\bar t)
\end{bmatrix}\right\rangle,
\quad f\in H^2.
\end{equation}

Thus we get 
\begin{theorem} The combination of the following two conditions
\begin{equation}\label{5sos}
\langle
(W^{-\frac 1 2}P_+W^{\frac 1 2})F, 
(W^{-\frac 1 2}P_+W^{\frac 1 2})F\rangle\le
Q
\langle
F, F\rangle
\end{equation}
with $F=W^{\frac 1 2}
\begin{bmatrix} f(t)\\
\bar t f(\bar t)
\end{bmatrix}$ and $f\in L^2_{s_+}\ominus H^2_{s_+}$, and
\begin{equation}\label{6sos}
\langle
(W^{\frac 1 2}P_+W^{-\frac 1 2})F, 
(W^{\frac 1 2}P_+W^{-\frac 1 2})F\rangle\le
Q
\langle
F, F\rangle
\end{equation}
with $F=W^{\frac 1 2}
\begin{bmatrix} f(t)\\
\bar t f(\bar t)
\end{bmatrix}$ and $f\in  H^2$ is equivalent  to the first (or the second) 
condition from  \cite{VYu}, Theorem 3.1.
\end{theorem}

\begin{proof} 
Relation \eqref{5sos} is a slight modification of
\eqref{1sos} and \eqref{6sos} of \eqref{4sos}, respectively.  Observe that if 
\eqref{6sos} holds with $f\in H^2$, then every $f\in H^2_{s_+}$ belongs to 
$H^2$. Therefore, in fact, \eqref{6sos}
has a perfect sense for $f\in H^2_{s_+}$.
\end{proof}

In any case, $W\in A_2$ is a sufficient condition for
\eqref{5sos}, \eqref{6sos}. Let us transform this matrix condition
into a scalar one.
\begin{lemma} $W$ is in $A_2$ if and only if
\begin{equation}\label{sina2}
\sup_{I}\frac{1}{|I|}\int_{I} 
\frac{|s_+-\Ms|^2+ (1-|\Ms|^2)}{1-|s_+|^2}
\,dm<\infty,
\end{equation}
where for an arc $I\subset \bbT$ we put
\begin{equation}
\Ms:=\frac{1}{|I|}\int_{I} s_+\,dm.
\end{equation}
\end{lemma}

\begin{proof}
By definition we have that there exists $Q>0$ such that
\begin{equation}\label{7sos}
\langle W^{-1}\rangle_{I}\le Q\langle W\rangle_{I}^{-1}
\end{equation}
for all $I\subset \bbT$.
Note that
\begin{equation*}
\langle W\rangle_{I}=\begin{bmatrix} 1&\overline{\Ms}\\
\Ms& 1\end{bmatrix}=
\begin{bmatrix} 1&0\\
\Ms& 1\end{bmatrix}
\begin{bmatrix} 1&0\\
0& 1-|\Ms|^2\end{bmatrix}
\begin{bmatrix} 1&\overline{\Ms}\\
0& 1\end{bmatrix}.
\end{equation*}
Therefore \eqref{7sos} is equivalent to
\begin{equation*}
\begin{bmatrix} 1&0\\
0& \sqrt{1-|\Ms|^2}\end{bmatrix}
\begin{bmatrix} 1&\overline{\Ms}\\
0& 1\end{bmatrix}
\langle W^{-1}\rangle_{I}
\begin{bmatrix} 1&0\\
\Ms& 1\end{bmatrix}
\begin{bmatrix} 1&0\\
0& \sqrt{1-|\Ms|^2}\end{bmatrix}
\le Q.
\end{equation*}
Since the matrix in the RHS is positive, its boundedness is equivalent
to the boundedness of its trace. The last condition 
with a small effort gives \eqref{sina2} and vice versa.
\end{proof}

\medskip\noindent
{\it Acknowledgments.}\ \ A part of this work was done during the last author's 
visit to CMI (Centre de Math\'ematique et Informatique) at University of 
Provence. He would like to thank the Department for the hospitality.

\end{document}